\documentclass[twocolumn,preprintnumbers,aps,prd,amsmath,amssymb]{revtex4}
\usepackage{amsmath,amssymb,amsfonts,epsfig,subfigure}
\usepackage{color}
\usepackage[utf8]{inputenc}
\usepackage{CJK}

\usepackage[T1]{fontenc}
\usepackage[colorlinks,linkcolor=blue,urlcolor=blue,citecolor=blue]{hyperref}

\usepackage{pifont}

\newcommand{\cmark}{\mbox{\ding{51}}}
\newcommand{\xmark}{\mbox{\ding{55}}}

\def \be {\begin{equation}}
\def \ee {\end{equation}}
\def \ba {\begin{aligned}}
\def \ea {\end{aligned}}
\def \bea {\begin{eqnarray}}
\def \eea {\end{eqnarray}}

\begin{document}
\begin{CJK*}{UTF8}{gbsn}

\title{Failure of Generalized Lennard-Jones Models for Charged AdS Black Hole Molecules in the Virial Expansion}

\thanks{C. H. Liu acknowledges support from the National Natural Science Foundation of China under Grant Nos.~12547167 and 12234019, and from the Postdoctoral Fellowship Program and the China Postdoctoral Science Foundation under Grant No.~GZC20261834.}

\author{Conghua Liu}
\affiliation{College of Physics, Jilin University, Changchun 130012, China}

\author{Jin Wang}
\email{jin.wang.1@stonybrook.edu}
\affiliation{Department of Chemistry and Department of Physics and Astronomy, State University of New York at Stony Brook, Stony Brook, New York 11794, USA}

\begin{abstract}
The microscopic interactions underlying the thermodynamic behavior of charged AdS black holes remain an open question. Motivated by the proposed description of black-hole molecules in terms of Lennard-Jones-type interactions, we examine whether a generalized $m\!-\!n$ Lennard--Jones pair potential can consistently reproduce the microscopic information encoded in both Ruppeiner geometry and the equation of state. We formulate the problem through the virial expansion, in which the interaction potential determines the virial coefficients via Mayer cluster integrals. The vanishing of the Ruppeiner scalar curvature fixes the characteristic equilibrium separation of the potential, while the second virial coefficient determines its energy scale. The third- and fourth-order virial coefficients then provide independent consistency constraints. By systematically scanning the repulsive and attractive exponents over different geometric prescriptions, temperatures, and black-hole charges, we find no parameter set for which the required third- and fourth-order conditions are simultaneously satisfied. In particular, the corresponding zero loci are either absent or, when both are present, do not intersect in parameter space. We further show that varying the charge at fixed $QT$ changes only the magnitudes of the relevant virial coefficients, without altering the locations of their zeros. These results demonstrate that, within the framework of classical pairwise-additive interactions and the adopted geometric prescription, the generalized Lennard-Jones models with exponents satisfying $30\geq m>n>3$ fail in the virial expansion for charged AdS black hole molecules.  Although such potentials may qualitatively capture the competition between short-range repulsion and longer-range attraction, they cannot serve as an exact microscopic pair potential that simultaneously reproduces the complete virial structure and satisfies the constraint imposed by the Ruppeiner geometry of the charged AdS black hole.

\end{abstract}

\maketitle

\end{CJK*}

\section{Introduction}
Since the groundbreaking work of Bekenstein and Hawking~\cite{AA,AB}, black holes have been widely regarded as thermodynamic systems endowed with thermodynamic quantities such as temperature and entropy. Like ordinary thermodynamic systems, black holes can also undergo phase transitions. For a Schwarzschild-AdS black hole, there exists a first-order phase transition between thermal radiation and a large black hole, which can be interpreted as a confinement/deconfinement phase transition in the context of the AdS/CFT correspondence~\cite{AC,AD}. For a charged AdS black hole, the small/large black-hole phase transition in the canonical ensemble, in which the charge is held fixed, is characterized by a first-order phase transition below the critical temperature and a second-order critical point at the endpoint of the coexistence line, analogous to the liquid/gas phase transition of a van der Waals fluid~\cite{AE,AF}. In the extended phase space, where the cosmological constant is interpreted as thermodynamic pressure, the analogy between charged AdS black holes and van der Waals fluids has been investigated more systematically~\cite{AG,AH,AI,AJ,AK,AL,AM,AN}. In particular, they possess analogous equations of state and share the same critical exponents~\cite{AJ,AK}.

Recently, the study of black hole thermodynamic phase transitions has been extended from thermodynamic analysis to the investigation of dynamical processes. In particular, the transitions between different black hole states have been described as stochastic processes governed by the Fokker-Planck equation in the generalized free energy landscape~\cite{ZZZZA,ZZZZB}. Within this framework, the mean first passage time and its fluctuations have been investigated to characterize the kinetics of black hole phase transitions. Subsequent studies have extended these investigations to various types of black holes~\cite{ZZZZC,ZZZZD,ZZZZE,ZZZZF,ZZZZG,ZZZZH,ZZZZI,ZZZZJ} and explored related theoretical approaches, including the Fokker-Planck equation in thermal potentials~\cite{ZZZZK}, path integral formulations of Langevin dynamics~\cite{ZZZZL,ZZZZM}, and Kramers escape rate theory~\cite{ZZZZN,ZZZZO,ZZZZP,ZZZZQ}. Furthermore, more rigorous theoretical approaches have been proposed to derive the generalized free energy~\cite{ZZZZM,ZZZZR,ZZZZS}.

The generalized free energy not only plays an important role in describing the dynamical processes of black hole phase transitions but also can be used to investigate black hole thermodynamic topology. In particular, a vector field can be constructed using the first derivative of the generalized free energy, and the corresponding winding number can be used to characterize the thermodynamic stability of black holes, with positive and negative winding numbers corresponding to stable and unstable black hole states, respectively~\cite{ZZZA}. This approach has subsequently been extended to various types of black holes~\cite{ZZZB,ZZZC,ZZZD,ZZZE,ZZZF,ZZZG,ZZZH,ZZZI,ZZZJ,ZZZK,ZZZL,ZZZM,ZZZN,ZZZO,ZZZP}.

The similarities between charged AdS black holes and van der Waals fluids are not limited to the first order phase transition below the critical temperature but also extend to the supercritical region beyond the critical point. Although no discontinuous phase transition occurs in the supercritical region due to the presence of a single thermodynamic phase, crossover lines can still be identified from thermodynamic and dynamical perspectives, commonly referred to as the Widom line and the Frenkel line, respectively. In particular, the Widom line has been investigated using the Lee-Yang theory of phase transitions through a complex analytic approach~\cite{ZKA,ZKB,ZKC,ZKD,ZKE}, higher derivatives of the Gibbs free energy~\cite{ZKF}, and the kinetics of black hole state transitions characterized by the autocorrelation time~\cite{ZKG}. In addition, the Frenkel line has been investigated using quasinormal modes~\cite{ZKF}. Furthermore, two supercritical crossover lines, denoted by $L^{\pm}$, have been introduced to divide the supercritical region into three distinct regimes corresponding to small black hole like, large black hole like, and intermediate supercritical regimes~\cite{ZKH,ZKI}.

These thermodynamic studies have not only deepened our understanding of black-hole physics but have also motivated extensive efforts to uncover the microscopic degrees of freedom underlying their macroscopic thermodynamic behavior. In this context, black holes have been proposed to possess a microscopic structure composed of so-called black-hole molecules, and a corresponding number density has been introduced to characterize their microscopic degrees of freedom~\cite{AO}. Furthermore, Ruppeiner geometry, a thermodynamic geometry whose line element measures the distance between two neighboring fluctuation states, can be used to probe the interactions among these microscopic degrees of freedom~\cite{AP,AQ,AR,AS,AT,AU,AV}. In particular, positive and negative Ruppeiner scalar curvatures are generally associated with repulsive and attractive interactions, respectively, whereas a vanishing scalar curvature corresponds to a noninteracting system. The Ruppeiner scalar curvature of a charged AdS black hole has been found to be positive in the small-volume regime and negative in the large-volume regime~\cite{AP,AT}, suggesting that repulsive and attractive microscopic interactions dominate in these two regimes, respectively.

Motivated by the analogy between charged AdS black holes and van der Waals fluids, it is natural to ask whether their microscopic interactions can be described by the same or similar interaction potentials. Since the Lennard-Jones \(12\text{--}6\) potential provides a simple effective description of short-range repulsion and longer-range attraction in ordinary fluids, it has been proposed as a phenomenological model for the effective interaction between charged AdS black-hole molecules~\cite{AW,AX,BD}. Such an interpretation, however, has so far remained largely qualitative, and the parameters of the proposed interaction potential have typically been determined within a mean-field approximation rather than through a systematic matching of the virial coefficients. Besides the mean-field approach, the virial expansion has also been employed to investigate the microscopic interactions among black hole molecules. In particular, Miao and Xu proposed an attractive interaction potential for the thermodynamically stable Schwarzschild-AdS (SAdS) black hole based on the microscopic information inferred from Ruppeiner geometry~\cite{KKKA}. The parameter of the proposed interaction potential was determined by matching the leading contribution to the second virial coefficient in the high-temperature limit, while the remaining contributions were interpreted as thermal corrections to the black-hole equation of state. Furthermore, the authors suggested that this approach could be extended to other black-hole systems through the calculation 
of higher-order virial coefficients. Along this line of research, we extend the virial-coefficient approach to charged AdS black holes and investigate whether the proposed microscopic interaction potential can simultaneously reproduce the second, third, and fourth virial coefficients required by the black-hole equation of state. Since the virial coefficients are determined by the cluster integrals of the intermolecular potential, such an analysis provides the order-by-order constraints on the proposed microscopic interaction potential and allows us to assess its applicability and limitations 
beyond the level of the second virial coefficient.

In this work, we generalize the Lennard-Jones \(12\text{--}6\) potential to the more general \(m\text{--}n\) form and examine whether this class of potentials can simultaneously reproduce the constraint inferred from Ruppeiner geometry and the complete virial-coefficient structure of the charged AdS black-hole equation of state, which contains nonvanishing second- and fourth-order virial terms, a vanishing third-order term, and no higher-order contributions. We find that these constraints cannot be satisfied simultaneously, even when the repulsive and attractive exponents are allowed to vary. The generalized Lennard-Jones pair potential may nevertheless serve as a qualitative effective model that captures the generic competition between short-range repulsion and longer-range attraction. It cannot, however, be regarded as an exact microscopic pair potential capable of reproducing the complete virial structure encoded in the equation of state of the charged AdS black hole and of satisfying the requirements of the Ruppeiner geometry. These results suggest that, within the framework of classical pairwise-additive interactions and the adopted geometric prescription, the generalized Lennard-Jones description of charged AdS black-hole molecules fails within the framework of the virial expansion.

%%%%%%%%%%%%%%%%%%%%%%%%%%%%%%%%%%%%%%%%%%%%%%%%%%%%%%%%%
%%%%%%%%%%%%%%%%%%%%%%%%%%%%%%%%%%%%%%%%%%%%%%%%%%%%%%%%%
\section{Equation of state of charged AdS black hole and the virial expansion}
The metric of the charged AdS black hole is written as~\cite{AJ,AK}:
\be
\ba
\label{eq:1}
ds^2=-F(r)dt^2+\frac{dr^2}{F(r)}+r^2d\Omega^2,
\ea
\ee
where $F(r)$ is given by  
\be
\ba
\label{eq:2}
F(r)=1-\frac{2M}{r}+\frac{Q^2}{r^2}+\frac{r^2}{L^2}.
\ea
\ee
Here $M$ is the mass, $Q$ is the charge, and $L$ is the AdS curvature radius associated with the negative cosmological constant $\Lambda$ by $L=\sqrt{{-3}/{\Lambda}}$.

The Hawking temperature is given by
\be
\ba
\label{eq:3}
T_H=\frac{F'(r)}{4\pi}|_{r=r_+}=\frac{1}{4\pi r_+}(1+\frac{3r_+^2}{L^2}-\frac{Q^2}{r_+^2}).
\ea
\ee
After interpreting the cosmological constant as the thermodynamic pressure $P=3/8\pi L^2$~\cite{AG,AH,AI,AJ,AK}, 
the temperature equation (\ref{eq:3}) can be rewritten as the equation of state
\be
\ba
\label{eq:4}
P=\frac{T}{v}-\frac{1}{2\pi v^2}+\frac{2Q^2}{\pi v^4},
\ea
\ee
where $v$ is the specific volume with the definition $v=2r_+$~\cite{AJ,AK}. 

In Ref.~\cite{AO}, the black hole is proposed to be composed of black hole molecules, with a molecular density given by $n=1/v$. Here, we 
investigate the interaction potentials between these black hole molecules from statistical physics. The virial expansion expresses the equation of state in the form of a power series of $1/v$ as:
\be
\ba
\label{eq:5}
\frac{P}{T}=\frac{1}{v}+\frac{\mathcal{V}_{2}(T)}{v^2}+\frac{\mathcal{V}_{3}(T)}{v^3}+\frac{\mathcal{V}_{4}(T)}{v^4}+\dots,
\ea
\ee
where $\mathcal{V}_{2}(T)$, $\mathcal{V}_{3}(T)$, and $\mathcal{V}_{4}(T)$ are called the second, third, and fourth virial coefficients, respectively. These virial coefficients are determined by the irreducible cluster integrals derived from the statistical mechanical partition function and can be calculated from the microscopic interaction potential, thereby establishing a connection between the microscopic interactions and the macroscopic equation of state. Below, we briefly outline the derivation.

We start from the canonical partition function for a system of $N$ molecules, which is given by $Z_N=(1/N!\lambda^{3N})\int \exp[-\beta U_N]\,d^{3N}r$, where $\beta=1/(k_B T)$, $N$ is the total number of molecules, and $\lambda$ is the thermal de Broglie wavelength. Assuming that the interactions between molecules are pairwise additive, the total interaction energy can then be expressed as
\(
U_N = \sum_{i<j} u_{ij}.
\) This interaction energy enters the partition function through the Boltzmann factor
\(
\exp[{-\beta U_N}] = \exp[{-\beta\sum_{i<j}u_{ij}}],
\) which is also the main focus of our derivation.

After introducing the Mayer function
\(
f_{ij} = e^{-\beta u_{ij}} - 1,
\) the Boltzmann factor can be rewritten as
\be
\ba
\label{eq:666}
\exp[{-\beta\sum_{i<j}u_{ij}}] = \prod_{i<j} e^{-\beta u_{ij}} = \prod_{i<j} (1 + f_{ij}).
\ea
\ee
Here, the term $1$ corresponds to the noninteracting contribution, while $f_{ij}$ represents the correction due to the interaction between molecules $i$ and $j$. Expanding the product $\prod_{i<j}(1+f_{ij})$ generates a sum of terms, each of which can be represented by a Mayer graph: each vertex denotes a molecule, and each edge $(ij)$ denotes a factor $f_{ij}$. This Mayer function expansion can be explicitly written as:
\be
\ba
\label{jdusbdb}
\prod_{i<j}(1+f_{ij})=1+\sum_{i<j}f_{ij}+\sum_{(i,j)<(k,l)}f_{ij}f_{kl}+\cdots.
\ea
\ee
The first term represents the contribution without intermolecular interactions, whereas the remaining terms account for contributions involving one or more Mayer functions.

Substituting Eq.~(\ref{jdusbdb}) into the canonical partition function and applying the standard Mayer cluster expansion, which involves classifying and summing the contributions from Mayer graphs, one can obtain~\cite{BA}
\be
\ba
\label{dbyyvxhssad}
\ln Z_N=\ln Z_N^{(0)}+N\sum_{l=1}^{\infty}\frac{\alpha_{l+1}}{(l+1)!V}(\frac{N}{V})^l
\ea
\ee
where $Z_N^{(0)}=V^N/(N!\lambda^{3N})$ represents the contribution of the first term $1$ in Eq.~(\ref{jdusbdb}), corresponding to the partition function of an ideal gas, and $\alpha_{l+1}$ denotes the irreducible cluster integral. Namely, not all Mayer graphs contribute independently to the virial coefficients. Disconnected graphs factorize into products of connected clusters, whereas reducible connected graphs can be generated from lower-order contributions. The genuinely $l+1$-body contribution is therefore represented by irreducible connected Mayer graphs containing $l+1$ vertices. Here, ``connected'' means that all vertices belong to a single cluster, while ``irreducible'' means that the graph contains no articulation vertex; equivalently, for a graph with at least three vertices, it remains connected after the removal of any single vertex and its incident edges. The corresponding irreducible cluster integral is~\cite{BA}
\be
\ba
\label{eq:667}
\alpha_{l+1}
=
\int d\mathbf r_0 d\mathbf r_1\cdots d\mathbf r_l
\sum_{G\in \mathcal B_{l+1}}
\prod_{(ij)\in E(G)} f_{ij}.
\ea
\ee
Here, \(\mathcal B_{l+1}\) denotes the set of all irreducible connected Mayer graphs with vertices
\(0,1,\ldots,l\), and \(E(G)\) is the set of edges of a given graph \(G\). For each graph \(G\), the product
\(
\prod_{(ij)\in E(G)} f_{ij}
\)
means multiplying the Mayer functions associated with all edges of that graph. The summation over
\(G\in\mathcal B_{l+1}\) then adds the contributions from all irreducible connected graphs containing
\(l+1\) vertices. Finally, the integration over
\(d\mathbf r_0 d\mathbf r_1\cdots d\mathbf r_l\) sums these graph weights over all molecular configurations.

Substituting Eq.~(\ref{dbyyvxhssad}) into the thermodynamic relation $P=k_BT(\partial \ln Z_N/\partial V)_{T,N}$ and setting $k_B=1$, one can then obtain the equation of state:
\be
\ba
\frac{P}{T}=\frac{N}{V}-\sum_{l=1}^{\infty}\frac{l}{(l+1)!V}\alpha_{l+1}(\frac{N}{V})^{l+1},
\ea
\ee
which reproduces Eq.~(\ref{eq:5}) upon identifying the specific volume as $v=V/N$. Thus, the $l+1$-th virial coefficient is then related to this irreducible cluster integral by~\cite{BA}
\be
\ba
\label{eq:668}
\mathcal{V}_{l+1}(T) = -\frac{l}{(l+1)!\,V}\,\alpha_{l+1}.
\ea
\ee
In summary, the intermolecular potential enters the equation of state through the Mayer functions $f_{ij}$. By summing products of these functions over irreducible connected Mayer graphs and integrating over the molecular coordinates, one obtains the corresponding irreducible cluster integrals, which determine the virial coefficients. On the other hand, the virial coefficients can also be determined directly from the black hole equation of state. Comparing the virial coefficients obtained through these two approaches thus provides a systematic way to test whether the proposed microscopic interaction potential is consistent with the macroscopic equation of state. More importantly, the virial coefficients at different orders impose distinct constraints on the same microscopic interaction potential. Reproducing a single virial coefficient is therefore insufficient to establish the consistency of the proposed potential with the equation of state. Instead, the proposed potential must simultaneously satisfy the constraints imposed by the virial coefficients at different orders. In this sense, the virial-coefficient approach provides a systematic framework for testing the proposed microscopic interaction potential beyond the qualitative characterization of microscopic interactions inferred from Ruppeiner thermodynamic geometry.

Before applying the above formalism to charged AdS black 
holes, it is useful to discuss its connection with the 
previous investigation of the microscopic interactions 
among SAdS black hole molecules in Ref.~\cite{KKKA}. 
The physical relationship between the two systems can be 
seen directly from their equations of state. Setting $Q=0$ 
in Eqs.~(\ref{eq:4}) recovers the equation of state of the SAdS 
black hole,
\[
P=\frac{T}{v}-\frac{1}{2\pi v^2}.
\]
For a nonvanishing charge $Q$, the coefficient of $v^{-2}$ 
remains unchanged, while an additional positive term 
proportional to $2Q^2/\pi v^4$ appears and the coefficient of 
$v^{-3}$ remains zero. From a phenomenological perspective, 
the negative $v^{-2}$ term can be associated with attractive 
microscopic interactions, whereas the additional positive 
$v^{-4}$ term in the charged AdS black-hole equation of state 
can be associated with repulsive interactions. This distinction 
provides a physical motivation for considering an interaction 
potential incorporating both attractive and repulsive 
contributions for charged AdS black holes, rather than the 
attractive potential adopted for SAdS black holes in 
Ref.~\cite{KKKA}.

\begin{figure*}[t]
\includegraphics[width=15cm]{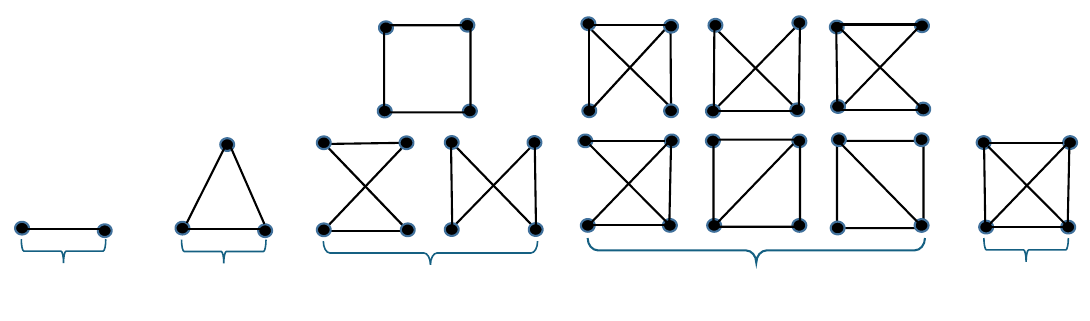}
\put(-406,13){\scalebox{1}{$\beta_2$}}
\put(-429.5,33){\scalebox{1}{$0$}}
\put(-380,33){\scalebox{1}{$1$}}
\put(-343,13){\scalebox{1}{$\beta_3$}}
\put(-368,33){\scalebox{1}{$2$}}
\put(-340,74){\scalebox{1}{$0$}}
\put(-317,33){\scalebox{1}{$1$}}
\put(-266,13){\scalebox{1}{$3\beta_4^{(1)}$}}
\put(-286,79){\scalebox{1}{$3$}}
\put(-238,79){\scalebox{1}{$2$}}
\put(-286,113){\scalebox{1}{$0$}}
\put(-238,113){\scalebox{1}{$1$}}
\put(-138,13){\scalebox{1}{$6\beta_4^{(2)}$}}
\put(-207,79){\scalebox{1}{$3$}}
\put(-159,79){\scalebox{1}{$2$}}
\put(-207,113){\scalebox{1}{$0$}}
\put(-159,113){\scalebox{1}{$1$}}
\put(-28,13){\scalebox{1}{$\beta_4^{(3)}$}
\put(-39,23){\scalebox{1}{$3$}}
\put(9,23){\scalebox{1}{$2$}}
\put(-39,53){\scalebox{1}{$0$}}
\put(9,53){\scalebox{1}{$1$}}}
\caption{Irreducible connected Mayer graphs contributing to $\beta_2$, $\beta_3$, and $\beta_4$. Each vertex represents a molecule, and each edge $(ij)$ represents a Mayer function $f_{ij}$. At fourth order, the three topologically distinct graph classes contribute with multiplicities $3$, $6$, and $1$, corresponding to $3\beta_4^{(1)}$, $6\beta_4^{(2)}$, and $\beta_4^{(3)}$, respectively, so that $\beta_4=3\beta_4^{(1)}+6\beta_4^{(2)}+\beta_4^{(3)}$.}
\label{Fig1}
\end{figure*}

Returning to the equation of state of RNAdS black hole, a comparison between Eqs.~(\ref{eq:4}) and (\ref{eq:5}) in the equation of state yields the virial coefficients: $\mathcal{V}_{2}(T)=-1/2\pi T$, $\mathcal{V}_{3}(T)=0$, and $\mathcal{V}_{4}(T)=2Q^2/\pi T$. Substituting these results into Eq. (\ref{eq:668}) gives the relations:
\be
\ba
\label{eq:8}
-\frac{1}{2\pi T}=&-\frac{1}{2}\int f_{01} d\mathbf{r}_1=-\frac{1}{2}\beta_{2},
\ea
\ee
\be
\ba
\label{eq:881}
0=&-\frac{1}{3}\int\int f_{01}f_{02}f_{12}d\mathbf{r}_1d\mathbf{r}_2=-\frac{1}{3}\beta_{3},
\ea
\ee
\be
\ba
\label{eq:882}
\frac{2Q^2}{\pi T}=&-\frac{1}{8}\{3\beta_4^{(1)}+6\beta_4^{(2)}+\beta_4^{(3)}\}=-\frac{1}{8} \beta_4,
\ea
\ee
where we have defined $\beta_{l+1}=\alpha_{l+1}/V$ for notational simplicity. The three fourth-order integrals ($\beta_4^{(1)}$, $\beta_4^{(2)}$, and $\beta_4^{(3)}$) correspond to the topologically distinct irreducible graphs for four molecules:
\be
\ba
\label{eq:9}
\beta_4^{(1)}=&\int\int\int f_{01}f_{12}f_{23}f_{30}d\mathbf{r}_1d\mathbf{r}_2d\mathbf{r}_3,\\
\beta_4^{(2)}=&\int\int\int f_{01}f_{12}f_{23}f_{30}f_{02}d\mathbf{r}_1d\mathbf{r}_2d\mathbf{r}_3,\\
\beta_4^{(3)}=&\int\int\int f_{01}f_{12}f_{23}f_{30}f_{13}f_{02}d\mathbf{r}_1d\mathbf{r}_2d\mathbf{r}_3.
\ea
\ee
The numerical prefactors (3, 6, and 1) preceding $\beta_4^{(i)}$ in Eq. (\ref{eq:882}) count the number of the same topological graph structure. The corresponding Mayer graphs for different $\beta$ are shown in Fig.~\ref{Fig1}.

%%%%%%%%%%%%%%%%%%%%%%%%%
%%%%%%%%%%%%%%%%%%%%%%%%%%%%%%%%%%%%%%%%%%%%%%%%%%%%%%%%%

To facilitate the numerical evaluation, Eqs.~(\ref{eq:8}--\ref{eq:882}) are recast in the following form, with the detailed derivations provided in the Supplemental Material~\cite{BC,CC}:
\be
\ba
\label{ljhdhnb}
\frac{1}{\pi T}=&\beta_2=\int_0^{\infty} 4\pi {r_1}^2 f(r_1) dr_1,\\
0=&\beta_3=8\pi^2\int_0^{\infty} dr_1 r_1 f(r_1)\int_0^{\infty} f(r_2){r_2}dr_2\\
&
\int_{|r_1-r_2|}^{|r_1+r_2|}f(r_{12}){r_{12}}dr_{12},\\
-\frac{16Q^2}{\pi T}=&\beta_4=3\beta_4^{(1)}+6\beta_4^{(2)}+\beta_4^{(3)}.
\ea
\ee
Here,
\be
\ba
\beta_4^{(1)}=&16\pi^3\int_0^{\infty} dr_2 \big[\int_0^{\infty} f(r_1){r_1}dr_1 \int_{|r_1-r_2|}^{|r_1+r_2|}f(r_{12})\\
&{r_{12}}dr_{12}\big]^2,\\
\beta_4^{(2)}=&16\pi^3\int_0^{\infty} dr_2 f(r_2)[\int_0^{\infty} f(r_1){r_1}dr_1 \int_{|r_1-r_2|}^{|r_1+r_2|}f(r_{12})\\
&{r_{12}}dr_{12}]^2,\\
\beta_4^{(3)}=&\sum_{n=0}^{\infty}{8\pi^3(2n+1)} \int_0^{\infty}f(r_{1}) dr_{1}\int_0^{\infty}f(r_{2})dr_{2}\int_0^{\infty}\\
&f(r_{3})dr_{3}\int_{|r_1-r_2|}^{|r_1+r_2|}P_n(\frac{r_1^2+r_2^2-r_{12}^2}{2r_1r_2})f(r_{12})r_{12}dr_{12}\\
&\int_{|r_2-r_3|}^{|r_2+r_3|}P_n(\frac{r_2^2+r_3^2-r_{23}^2}{2r_2r_3})f(r_{23})r_{23}dr_{23}\int_{|r_1-r_3|}^{|r_1+r_3|}\\
&P_n(\frac{r_1^2+r_3^2-r_{13}^2}{2r_1r_3})f(r_{13})r_{13}dr_{13}.
\ea
\ee

In the numerical evaluation, the expansion in Legendre polynomials $P_n$ is truncated at $n=10$, and further increasing the truncation order does not significantly affect the resulting virial coefficients. The radial integrations entering the cluster integrals are performed over the full range of intermolecular separations, $0<r<\infty$, without imposing an explicit hard-core exclusion radius. Instead, the effective short-range exclusion is incorporated directly through the chosen interaction potential and hence through the Mayer function $f_{ij}=e^{-\beta u_{ij}}-1$, with the associated length scale determined by the parameters of the potential.

\section{Generalized Lennard-Jones potential and the occupancy fraction}
As discussed in the Introduction, the Lennard-Jones potential has been proposed as a candidate model for describing the effective interactions between charged AdS black-hole molecules. Here, we examine whether it can simultaneously satisfy the constraint inferred from Ruppeiner geometry and reproduce the complete virial-coefficient structure of the charged AdS black-hole equation of state, which terminates at fourth order. To this end, we consider the following generalized \(m\text{--}n\) Lennard-Jones potential:
\be
\ba
\label{eq:29}
u(r)=\frac{A}{r^m}-\frac{B}{r^n},
\ea
\ee
where \(A,B>0\) and \(m>n\). Equivalently, this potential can be parametrized as
\be
\ba
\label{eq:30}
u(r)=(\frac{m^m}{n^n})^{\frac{1}{m-n}}\frac{E}{m-n}[(\frac{d}{r})^m-(\frac{d}{r})^n],
\ea
\ee
where $r=d$ is the zero point of $u(r)$, \(E>0\) is the depth of the potential well, and the exponents \(m\) and \(n\) control the steepness of the repulsive and attractive branches, respectively. The minimum occurs at \(
r_{\min}=\left(\frac{m}{n}\right)^{\frac{1}{m-n}}d,
\)
where \(u(r_{\min})=-E\). For later convenience, we define $b=\frac{1}{m-n}(\frac{m^m}{n^n})^{\frac{1}{m-n}}$, and thus $u(r)=bE[(\frac{d}{r})^m-(\frac{d}{r})^n]$.

To determine the parameters of the Lennard-Jones potential, we first use
the microscopic information encoded in the Ruppeiner geometry~\cite{AP,AQ,AR,AS,AT,AU,AV}. In this
geometric description, the sign of the scalar curvature characterizes the
dominant microscopic interaction, whereas \(R=0\) marks the crossover at which the attractive and repulsive effects balance. For the charged AdS black hole,
Ref.~\cite{AP} showed that the scalar curvature $R=0$ when the specific volume of the charged AdS black hole molecule satisfies $v=2\sqrt{2}Q$. Furthermore, it should be emphasized that the association of positive and negative Ruppeiner thermodynamic curvature with repulsive and attractive microscopic interactions provides a qualitative characterization rather than an exact quantitative description. In the present work, we adopt this phenomenological interpretation as a constraint on the generalized LJ potential. Therefore, the resulting description of microscopic interactions among charged AdS black hole molecules should also be regarded as phenomenological.

To relate the specific volume at which the Ruppeiner curvature vanishes to the equilibrium separation of the generalized LJ potential, we introduce a characteristic sphere whose diameter is equal to the equilibrium separation, namely, the position of the potential minimum, $r_{\min}=\left(\frac{m}{n}\right)^{\frac{1}{m-n}}d$. Different spatial arrangements of these characteristic spheres generally lead to different correspondences between the specific volume and the equilibrium separation. To characterize such differences, we introduce an occupancy fraction $\eta$, which is defined as the ratio of the volume of the characteristic sphere to the specific volume. In this sense, $\eta$ serves as a geometric parameter that establishes a phenomenological connection between the characteristic volume obtained from Ruppeiner thermodynamic geometry and the equilibrium separation of the generalized LJ potential. Since the vanishing Ruppeiner curvature occurs at $v=2\sqrt{2}Q$, this correspondence leads to
\be
\ba
\label{eq:31}
\eta \times 2\sqrt{2}Q = \frac{4\pi}{3}
\left[
\frac{
\left(\frac{m}{n}\right)^{\frac{1}{m-n}} d
}{2}
\right]^3 .
\ea
\ee

The three representative values of $\eta$ considered in this work are motivated by previous studies, although the occupancy fraction was not explicitly introduced therein. Specifically, Ref.~\cite{AW} identifies the specific volume with $r_{\min}^3$, which corresponds to a simple cubic arrangement of the characteristic spheres. The corresponding occupancy fraction can be determined directly from geometric considerations. In a simple cubic arrangement, eight characteristic spheres are located at the eight vertices of a cube, with each sphere contributing one eighth of its volume to the cube. Thus, each cube effectively contains $8\times 1/8=1$ characteristic sphere. Moreover, the side length of the cube is equal to the diameter of the characteristic sphere, $r_{\min}$. Therefore, the occupancy fraction is given by $\eta=[4\pi(r_{\min}/2)^3/3]/r_{\min}^3=\pi/6$. Similarly, the prescription adopted in Ref.~\cite{BD} corresponds to the closest packing of the characteristic spheres, for which the occupancy fraction is $\eta=\sqrt{2}\pi/6$. By contrast, Ref.~\cite{AX} identifies the entire specific volume with the volume of the characteristic sphere, without accounting for any interstitial volume, which corresponds to $\eta=1$ in our framework. Therefore, these three representative values of $\eta$ reflect different prescriptions adopted in previous studies for relating the specific volume to the equilibrium separation of the LJ potential. In the following, we consider these three representative values, $\eta=\sqrt{2}\pi/6$, $\pi/6$, and $1$, to examine whether the generalized LJ potential can satisfy the required constraints under these different prescriptions. Then, we also extend our analysis to a wide range of occupancy fractions between $0$ and $1$ to examine whether the required constraints can be satisfied for other choices of $\eta$.

There are four parameters as $E$, $d$, $m$ and $n$ of the Lennard-Jones potential in Eq.~(\ref{eq:30}). Combining the Ruppeiner geometry (Eq.~(\ref{eq:31})) and the virial coefficients from the second to fourth order (Eq.~(\ref{eq:8}-\ref{eq:882})), we can obtain four equations. In principle, the parameters can therefore be determined by solving these equations, provided that a physically admissible solution exists.

However, in contrast to the direct calculation of virial coefficients for a prescribed interaction potential, which can be carried out numerically by substituting the potential into Eq.~(\ref{ljhdhnb}), reconstructing the interaction potential from a given set of virial coefficients is a substantially more challenging inverse problem because it requires solving a coupled set of nonlinear, multidimensional integral equations. To make this problem computationally tractable, we adopt the following parameter-scanning procedure.

We first note that the parameters $m$ and $n$ are not arbitrary, but are subject to certain constraints. We consider the calculation of $\beta_2$ as an example:
\be
\ba
\label{eq:32}
-\frac{1}{2\pi T}=&-\frac{1}{2}\int f_{01} d\vec{r_1}\\
=&-2\pi \int_0^{\infty} r^2 \{\exp[-\frac{bE}{T}((\frac{d}{r})^m-(\frac{d}{r})^n)]-1\}dr.
\ea
\ee
In the short-distance limit, \(u(r)\to+\infty\), and hence the Mayer function approaches \(f(r)=e^{-\beta u(r)} - 1\to-1\). Consequently, the radial integrand satisfies \(r^2f(r)\to-r^2\), so that the integral is convergent at the lower limit. In the large-distance limit, \(u(r)\to0\), and the Mayer function can be expanded as
\be
\ba
f(r) &= -\frac{u(r)}{T}+O\left(u^2\right) \\
&\simeq -\frac{bE}{T}
\left[
\left(\frac{d}{r}\right)^m
-\left(\frac{d}{r}\right)^n\right].
\ea
\ee

The large-distance contribution to the second virial coefficient therefore behaves as
\be
\ba
-2\pi\int_R^\infty r^2f(r)\,dr
\simeq
\frac{2\pi bE}{T}
\int_R^\infty
\left(
\frac{d^{\,m}}{r^{\,m-2}}
-
\frac{d^{\,n}}{r^{\,n-2}}
\right)dr,
\ea
\ee
where \(R\) is sufficiently large for the asymptotic expansion to be valid. Since \(m>n\), the slowly decaying attractive term proportional to \(r^{-n}\) determines the convergence of the integral. The tail integral behaves as \(\int^\infty r^{2-n}dr\), which is convergent only when \(n>3\). Thus, for a conventional generalized Lennard--Jones potential in three spatial dimensions, the exponents should satisfy
\be
\ba
m>n>3 .
\ea
\ee

Then, we consider \(n=3.1\) together with integer values of \(n\) ranging from \(4\) to \(30\). The upper bound of 30 is chosen to balance the coverage of the parameter space and the computational cost. This range includes the conventional Lennard-Jones 12-6 potential and allows us to examine a broad variety of interaction potentials with different repulsive and attractive exponents. Since the numerical evaluation of the multidimensional integrals associated with the third and fourth virial coefficients must be repeated for numerous parameter combinations at different temperatures and occupancy fractions, extending the scan to substantially larger exponents would considerably increase the computational cost. It should be emphasized that the upper bound of 30 is a practical numerical choice rather than a theoretical restriction. Accordingly, our numerical conclusions are limited to the parameter range investigated. For each value of \(n\), we scan over integer values of \(m\) satisfying $n<m\leq 30$. For each trial pair \((m,n)\), we substitute these values into the constraint derived from the Ruppeiner geometry, Eq.~(\ref{eq:31}), to determine \(d\). With \(m\), \(n\), and \(d\) fixed, the depth of the potential well \(E\) is then determined by requiring the second virial coefficient to satisfy Eq.~(\ref{eq:8}). This procedure yields a complete trial parameter set \((m,n,d,E)\).

Using each trial parameter set, we subsequently calculate the third and fourth virial coefficients, \(\beta_3\) and \(\beta_4\). According to Eqs.~(\ref{eq:881}) and~(\ref{eq:882}), an admissible solution must satisfy $\beta_3=0$ and \(\beta_4+16Q^2/(\pi T)=0\). A parameter set satisfying both conditions is therefore regarded as a solution that simultaneously fulfills the constraint inferred from the Ruppeiner geometry and reproduces the complete virial-coefficient structure of the charged AdS black-hole equation of state. If the initial scan identifies a candidate region in which both conditions are satisfied or approximately satisfied, we further refine the search within that region by scanning both \(m\) and \(n\) in increments of \(0.1\).

\section{Numerical results}
We now present the numerical results for the third- and fourth-order virial coefficients. For each trial pair \((m,n)\), Eqs.~(\ref{eq:31}) and~(\ref{eq:8}) determine \(d\) and \(E\), respectively, and are therefore always satisfied. It remains to examine whether the two conditions given in Eqs.~(\ref{eq:881}) and (\ref{eq:882}), namely, \(\beta_3=0\) and \(\beta_4+16Q^2/(\pi T)=0\), can be satisfied simultaneously.

\subsection{Occupancy fraction dependence}
%%%%%%%%%%%%%%%%%%%%%%%%%
\begin{figure}[t]
\includegraphics[width=8.5cm]{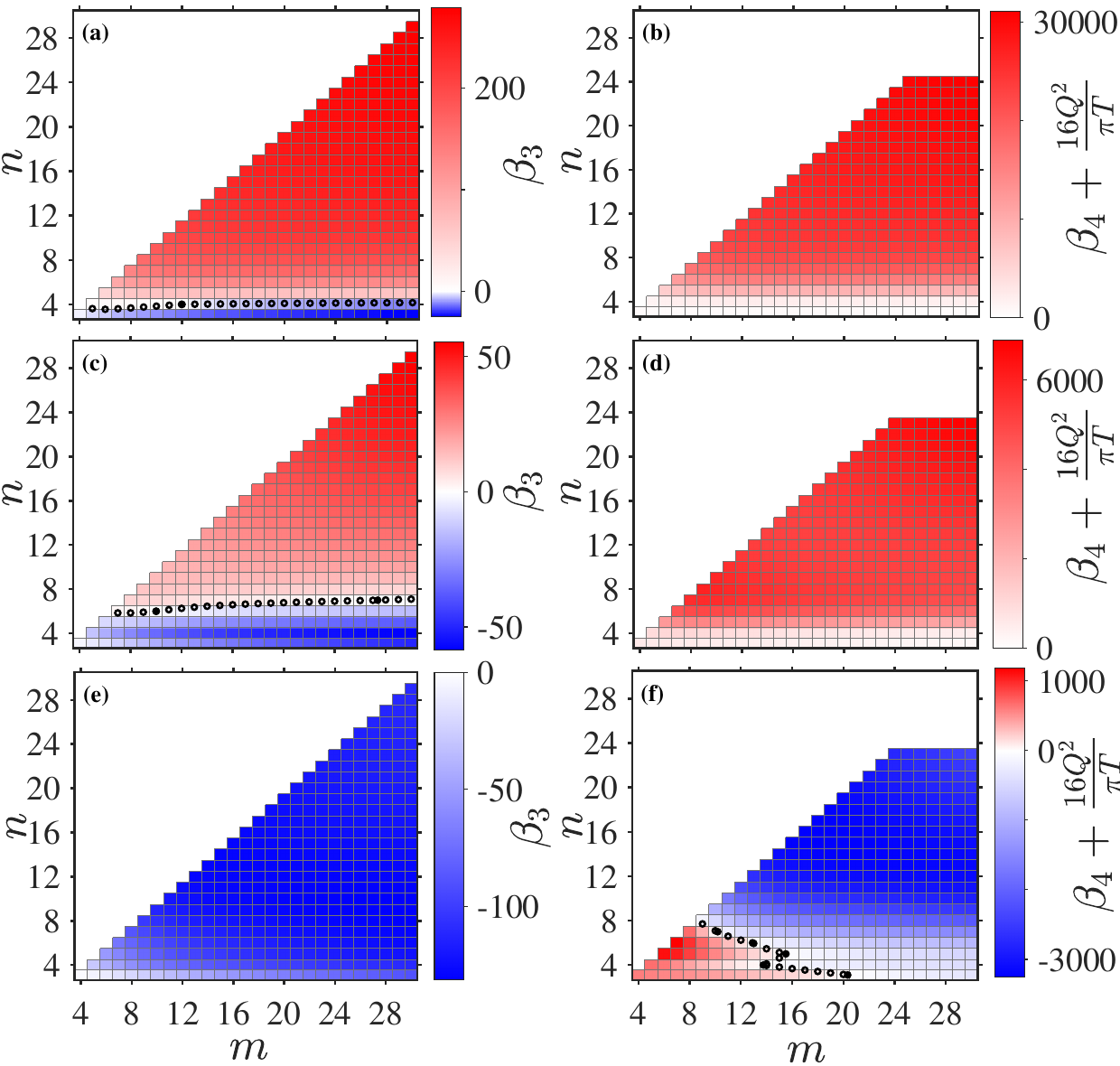}
\caption{The numerical results for $\beta_3$ and $\beta_4+{16Q^2}/{\pi T}$ at $Q=1$ and $T=0.2T_c$. The black filled circles denote the zeros obtained by fixing \(n\) and varying \(m\), whereas the black open circles denote those obtained by fixing \(m\) and varying \(n\). (a,b): The occupancy fraction is ${\pi}/{6}$, corresponding to the simple-cubic prescription. (c,d): The occupancy fraction is $\sqrt{2}\pi/6$, corresponding to the closest-packed prescription. (e,f): The occupancy fraction is $1$, corresponding to the prescription in which the entire specific volume is occupied.}
\label{Fig2}
\end{figure}
%%%%%%%%%%%%%%%%%%%%%%%%%%%%%%%%%%%%%%%%%%%%%%%%%%%

We take \(Q=1\) and \(T=0.2T_c\), where the critical temperature is \(T_c=\sqrt{6}/(18\pi Q)\). Figure~\ref{Fig2} shows the resulting values of \(\beta_3\) and \(\beta_4+16Q^2/(\pi T)\) over the scanned \((m,n)\) parameter space for different occupancy fractions. The black filled circles denote the zeros obtained by fixing \(n\) and varying \(m\), whereas the black open circles denote those obtained by fixing \(m\) and varying \(n\). These zero points allow us to examine whether the two required conditions can be satisfied simultaneously at the same point in the \((m,n)\) plane.

For the simple-cubic prescription with occupancy fraction \(\eta=\pi/6\), there exist parameter values for which \(\beta_3=0\), whereas \(\beta_4+16Q^2/(\pi T)\) remains nonzero throughout the scanned \((m,n)\) region, as shown in Fig.~\ref{Fig2}(a,b). The same behavior is observed in Fig.~\ref{Fig2}(c,d) for the closest-packed prescription with the larger occupancy fraction \(\eta=\sqrt{2}\pi/6\). By contrast, for \(\eta=1\), zero points of \(\beta_4+16Q^2/(\pi T)\) appear, while \(\beta_3\) remains nonzero throughout the scanned parameter region, as shown in Fig.~\ref{Fig2}(e,f). Therefore, for all three representative occupancy fractions, no trial parameter set \((m,n,d,E)\) simultaneously satisfies all the imposed constraints.

%%%%%%%%%%%%%%%%%%%%%%%%%
\begin{figure}[t]
\includegraphics[width=8.5cm]{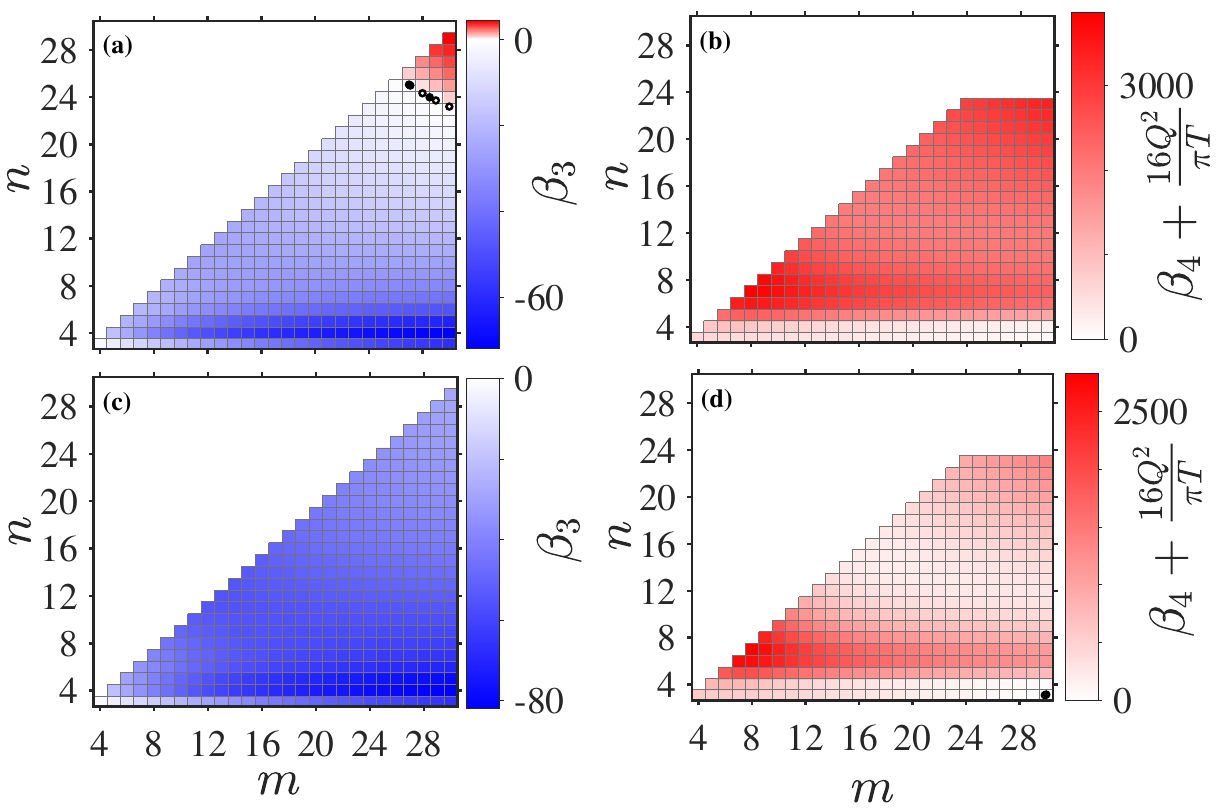}
\caption{The numerical results at the threshold values of the occupancy fraction where the zero-point structure changes as \(\eta\) is increased from \(\sqrt{2}\pi/6\) to \(1\). Panels (a) and (b) correspond to \(\eta=0.81\), beyond which the zero locus of \(\beta_3\) disappears. Panels (c) and (d) correspond to \(\eta=0.86\), at which zeros of \(\beta_4+16Q^2/(\pi T)\) first appear.}
\label{Fig3}
\end{figure}
%%%%%%%%%%%%%%%%%%%%%%%%%%%%%%%%%%%%%%%%%%%%%%%%%%%

The contrasting behaviors of \(\beta_3\) and \(\beta_4+16Q^2/(\pi T)\) as the occupancy fraction is varied naturally motivate us to further examine intermediate values of \(\eta\) between \(\sqrt{2}\pi/6\) and \(1\). Such intermediate values correspond to different geometric prescriptions for the characteristic sphere associated with the specific volume. The numerical results in Fig.~\ref{Fig3} show that the above conclusion remains unchanged for these intermediate values of $\eta$. As \(\eta\) is increased from \(\sqrt{2}\pi/6\), the zero locus of \(\beta_3\) gradually disappears, whereas the zero locus of \(\beta_4+16Q^2/(\pi T)\) emerges at larger \(\eta\). However, the ranges of $\eta$ over which zeros of $\beta_3$ and $\beta_4+16Q^2/(\pi T)$ occur do not overlap. Specifically, no zeros of $\beta_3$ are found for $\eta\geq 0.82$, whereas zeros of $\beta_4+16Q^2/(\pi T)$ appear only for $\eta\geq 0.86$. Consequently, the two types of zero regions never coexist for the same occupancy fraction, and the conditions \(\beta_3=0\) and \(\beta_4+16Q^2/(\pi T)=0\) cannot be satisfied simultaneously within the generalized Lennard--Jones model under these geometric prescriptions.

%%%%%%%%%%%%%%%%%%%%%%%%%
\begin{figure}[t]
\includegraphics[width=8.5cm]{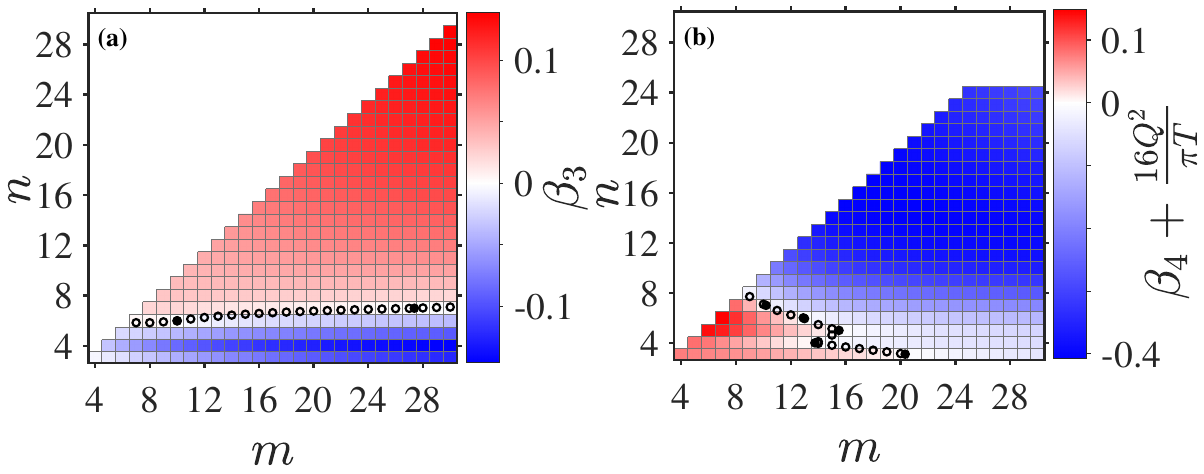}
\caption{The numerical results for $\beta_3$ and $\beta_4+{16Q^2}/{\pi T}$ at $Q=0.05$ and $T=0.2T_c$. (a): The occupancy fraction is $\sqrt{2}\pi/6$, corresponding to the closest-packed prescription. (b) The occupancy fraction is $1$, corresponding to the prescription in which the entire specific volume is occupied. Since the product $QT$ is identical to that used in Fig.~\ref{Fig2}, the locations of the zero points in the (m,n) parameter space coincide with those in Figs.~\ref{Fig2}(c) and \ref{Fig2}(f), respectively.}
\label{Fig4}
\end{figure}
%%%%%%%%%%%%%%%%%%%%%%%%%%%%%%%%%%%%%%%%%%%%%%%%%%%

\subsection{Charge dependence}
As shown above, for $Q=1$ and $T=0.2T_c$, the generalized Lennard--Jones potential cannot simultaneously satisfy the constraints inferred from the Ruppeiner geometry and reproduce the virial coefficients derived from the black-hole equation of state. We now examine whether these conditions can be satisfied for other values of the charge and temperature.

%%%%%%%%%%%%%%%%%%%%%%%%%
\begin{figure}[t]
\includegraphics[width=8.5cm]{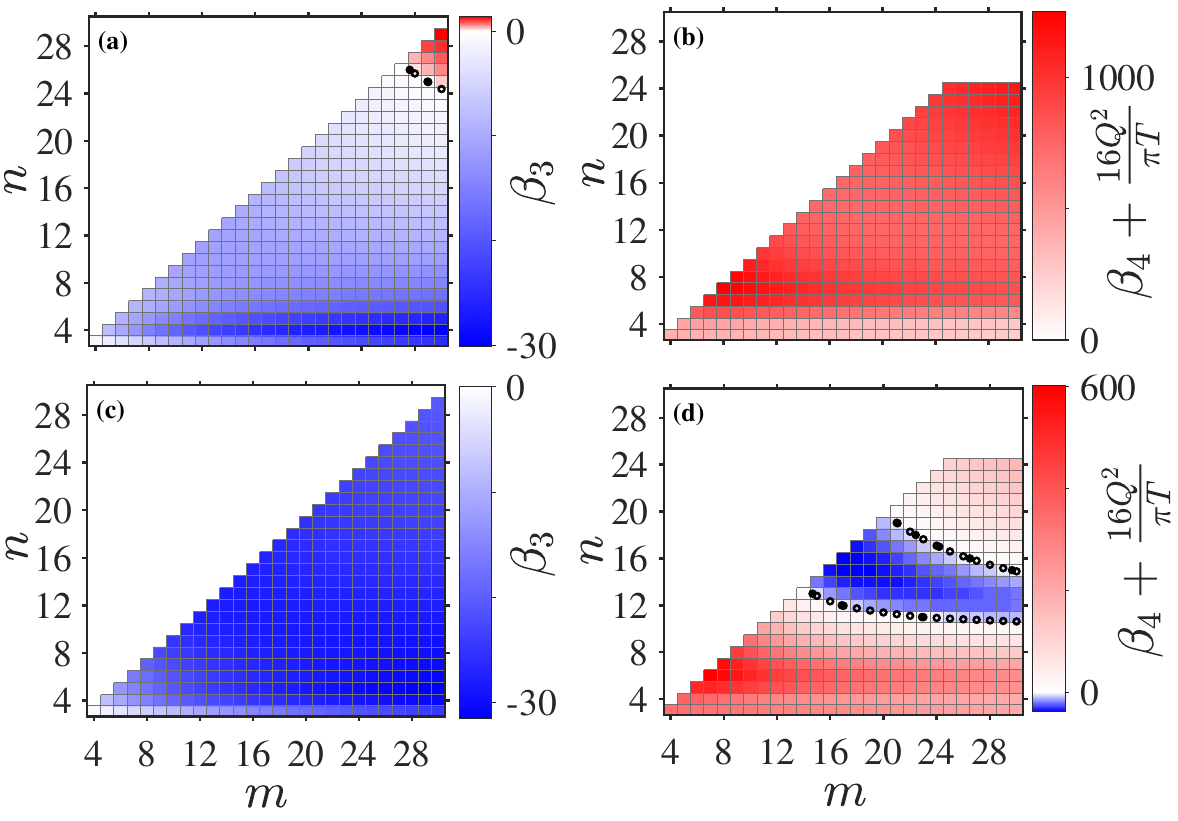}
\caption{The numerical results for $\beta_3$ and $\beta_4+16Q^2/(\pi T)$ at $Q=1$ for the simple-cubic prescription with $\eta=\pi/6$ at different temperatures. (a,b) $T=0.31T_c$. Above this temperature, the zeros of $\beta_3$ disappear. (c,d) $T=0.35T_c$. At this temperature, zeros of $\beta_4+16Q^2/(\pi T)$ first appear and persist at higher temperatures.}
\label{Fig5}
\end{figure}
%%%%%%%%%%%%%%%%%%%%%%%%%%%%%%%%%%%%%%%%%%%%%%%%%%%

First, we examine the dependence on the charge $Q$. Our numerical calculations show that, when the product $QT$ is kept fixed, the magnitudes of $\beta_3$ and $\beta_4+16Q^2/(\pi T)$ vary with $Q$, whereas their sign distributions over the scanned $(m,n)$ parameter space remain unchanged. Consequently, the locations of their zeros are also unchanged.

This behavior can be understood from the scaling properties of the virial coefficients. For a fixed occupancy fraction, the constraint from the Ruppeiner geometry in Eq.~(\ref{eq:31}) gives $d^3\propto Q$. Introducing the dimensionless radial variable $x=r/d$, the second virial coefficient in Eq.~(\ref{ljhdhnb}) can be written in the form
\be
\beta_2=d^3 F_2(m,n,E/T),
\ee
where $F_2$ is dimensionless. At fixed $QT$, both $d^3$ and the required value $\beta_2=1/(\pi T)$ scale linearly with $Q$. Therefore, the resulting dimensionless ratio $E/T$, where $E$ is determined from the second virial coefficient in our procedure, is independent of $Q$. It then follows that
\be
\beta_3=d^6 F_3(m,n,E/T)\propto Q^2,
\ee
while
\be
\beta_4=d^9 F_4(m,n,E/T)\propto Q^3.
\ee
Since the required third virial coefficient is $\beta_3=0$, the overall positive factor $Q^2$ does not affect the sign distribution or the locations of the zeros of $\beta_3$. The required fourth virial coefficient is $\beta_4=-16Q^2/(\pi T)$, which also scales as $Q^3$ at fixed $QT$. Therefore, $\beta_4+16Q^2/(\pi T)$ likewise acquires only an overall positive $Q^3$ factor. Hence, varying $Q$ at fixed $QT$ changes the magnitudes of these quantities but not their signs or the locations of their zeros in the $(m,n)$ parameter space. For the case considered here, $T=0.2T_c$, with $T_c={\sqrt{6}}/{18\pi Q}$, the product $QT$ is $QT=0.2\sqrt{6}/(18\pi)$, which is independent of $Q$. To verify this scaling behavior numerically, we change the charge from $Q=1$ to $Q=0.05$ while keeping $T/T_c=0.2$. We examine $\beta_3$ for the closest-packed prescription with $\eta=\sqrt{2}\pi/6$, and $\beta_4+16Q^2/(\pi T)$ for the prescription in which the entire specific volume is occupied, with $\eta=1$, as shown in Fig.~\ref{Fig4}. Consistent with the above scaling argument, the locations of the corresponding zeros remain unchanged.

%%%%%%%%%%%%%%%%%%%%%%%%%
\begin{figure}[t]
\includegraphics[width=8.5cm]{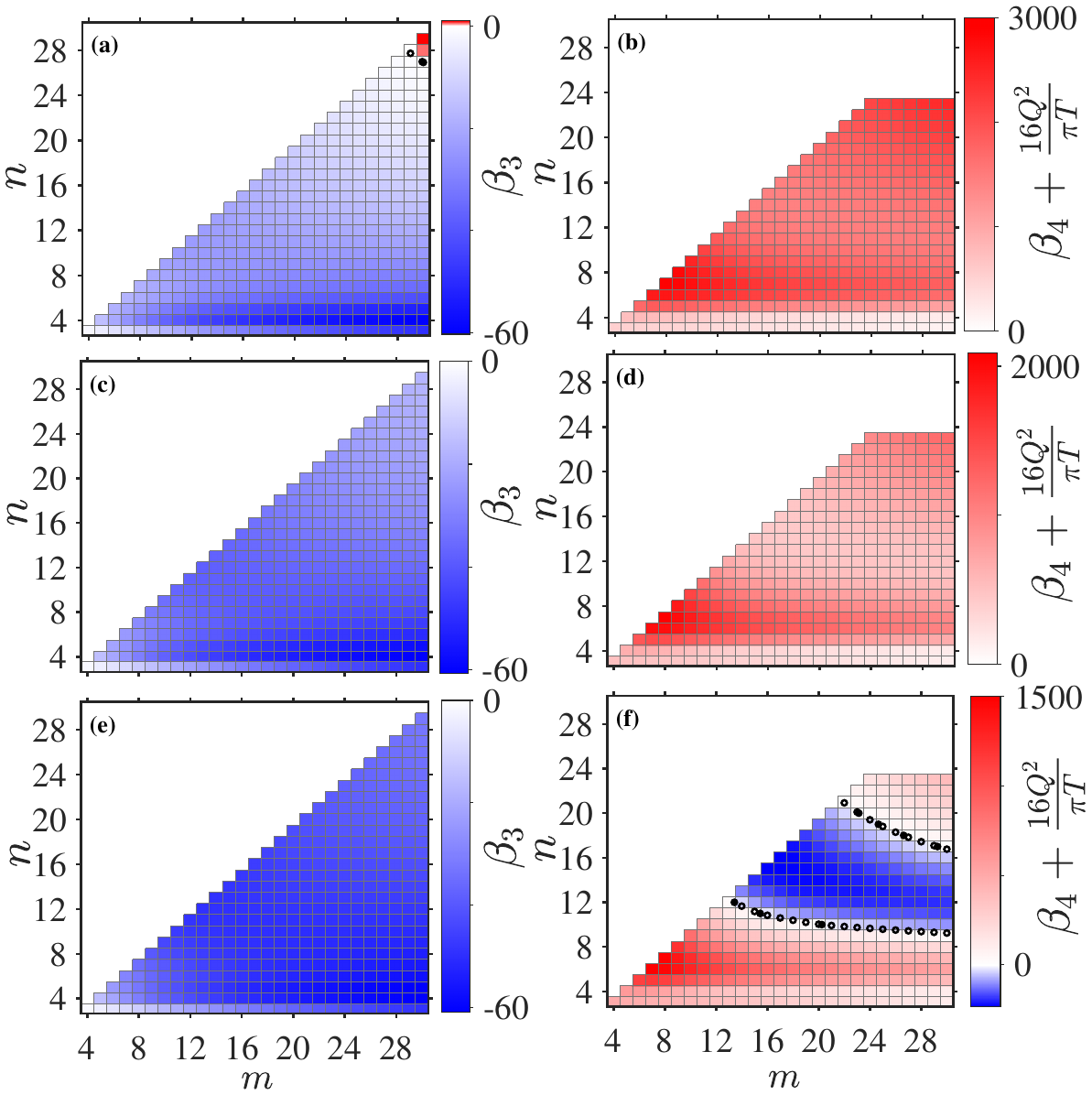}
\caption{The numerical results for $\beta_3$ and $\beta_4+16Q^2/(\pi T)$ at $Q=1$ for the closest-packed prescription with $\eta=\sqrt{2}\pi/6$ at different temperatures. (a,b) $T=0.22T_c$. Above this temperature, the zeros of $\beta_3$ disappear. (c,d) $T=0.23T_c$. At this temperature, neither $\beta_3$ nor $\beta_4+16Q^2/(\pi T)$ has zeros in the scanned $(m,n)$ parameter region. (e,f) $T=0.24T_c$. At this temperature, zeros of $\beta_4+16Q^2/(\pi T)$ first appear and persist at higher temperatures.}
\label{Fig6}
\end{figure}
%%%%%%%%%%%%%%%%%%%%%%%%%%%%%%%%%%%%%%%%%%%%%%%%%%%
\subsection{Temperature dependence}
As discussed above, varying the charge $Q$ at fixed $QT$ does not change the locations of the zeros. Therefore, for a fixed occupancy fraction, the temperature is the only remaining thermodynamic parameter whose effect needs to be examined. We now investigate whether, for the three characteristic occupancy fractions $\eta=\pi/6$, $\sqrt{2}\pi/6$, and $1$, the required constraints can be simultaneously satisfied at other temperatures.

%%%%%%%%%%%%%%%%%%%%%%%%%
\begin{figure}[t]
\includegraphics[width=8.5cm]{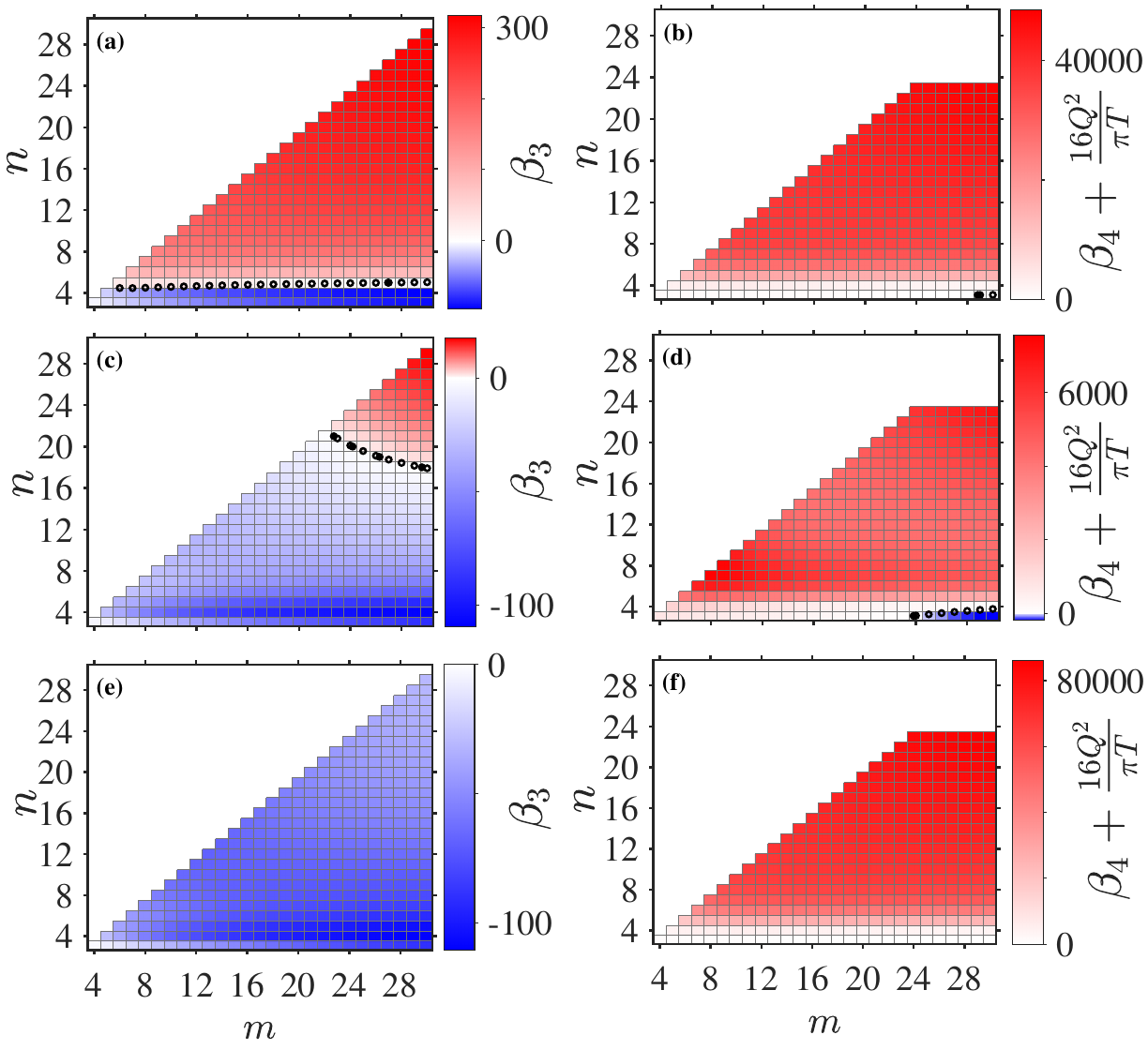}
\caption{Numerical results for $\beta_3$ and $\beta_4+16Q^2/(\pi T)$ with $Q=1$ and $\eta=1$ at different temperatures. 
(a,b) $T=0.13T_c$.  
(c,d) $T=0.16T_c$. 
(e) $T=0.17T_c$. (f) $T=0.12T_c$. Above the temperature $T=0.16T_c$, the zeros of $\beta_3$ vanish. Below the temperature $T=0.13T_c$, the zeros of $\beta_4+16Q^2/(\pi T)$ vanish.}
\label{Fig7}
\end{figure}
%%%%%%%%%%%%%%%%%%%%%%%%%%%%%%%%%%%%%%%%%%%%%%%%%%%

We first consider the simple-cubic case with occupancy fraction $\eta=\pi/6$ and vary the temperature to determine whether the required potential exists~\cite{AW}. We have examined a wide range of temperatures between $0.1T_c$ and $0.9T_c$, while Fig.~\ref{Fig5} presents only the representative temperatures near the thresholds at which the zero-point structure changes. Specifically, above the temperature $T=0.31T_c$, the zeros of $\beta_3$ disappear, while at temperature $T=0.35T_c$, zeros of $\beta_4+16Q^2/(\pi T)$ generally appear and do not occur at lower temperatures. Hence, the temperature ranges in which the two zero conditions occur do not overlap, and no temperature at this occupancy fraction allows the two constraints to be satisfied simultaneously.

Next, we consider the closest-packed prescription with occupancy fraction $\eta=\sqrt{2}\pi/6$ and study the temperature dependence~\cite{BD}. The results are shown in Fig.~\ref{Fig6}, where the representative temperatures near the thresholds at which the zero-point structure changes are presented. Specifically, within the scanned $(m,n)$ parameter region, the zeros of $\beta_3$ disappear for $T>0.22T_c$, whereas zeros of $\beta_4+16Q^2/(\pi T)$ appear only for $T\geq0.24T_c$. At the intermediate temperature $T=0.23T_c$, neither quantity has zeros. Therefore, the temperature ranges over which the two zero conditions occur do not overlap within the scanned parameter region, and no common zero of $\beta_3$ and $\beta_4+16Q^2/(\pi T)$ is found for the closest-packed prescription.

Finally, we examine the case in which the entire specific volume is occupied by the characteristic sphere, corresponding to an occupancy fraction of \(1\)~\cite{AX}, as shown in Fig.~\ref{Fig7}. In this case, for temperatures above \(T=0.16T_c\), no zeros of \(\beta_3\) exist, whereas zeros of \(\beta_4+16Q^2/(\pi T)\) can still occur. For temperatures below \(T=0.13T_c\), the zeros of \(\beta_4+16Q^2/(\pi T)\) vanish. For temperatures between these two temperatures, although zeros of both quantities may exist, they do not occur at the same \((m,n)\) values. Thus, this occupancy fraction also does not yield the required potential.

It is then natural to ask whether the required potential can be found for other occupancy fractions over the entire temperature range considered. We therefore extend the analysis to additional values of $\eta$ between $0$ and $1$ in increments of $0.1$, with the corresponding numerical results presented in the Supplemental Material~\cite{CC}. The conclusion remains unchanged, as summarized in Table~\ref{tab1}. For the sampled values in the range $0.01 \leq \eta \leq 0.02$, neither $\beta_3$ nor $\beta_4+16Q^2/(\pi T)$ has any zeros. For the sampled values in the range $0.03\leq\eta\leq0.36$, zeros of $\beta_3$ occur, whereas no zeros of $\beta_4+16Q^2/(\pi T)$ are found. For $0.37\leq\eta\leq0.88$, zeros of both $\beta_3$ and $\beta_4+16Q^2/(\pi T)$ occur, but in distinct temperature ranges; hence, the two zero conditions cannot be satisfied at the same temperature. For $0.89\leq\eta\leq1$, zeros of both quantities may occur at the same temperature; however, their zero loci do not intersect in the scanned $(m,n)$ parameter plane. Consequently, no common zero of $\beta_3$ and $\beta_4+16Q^2/(\pi T)$ is found for any of the additional occupancy fractions considered.

\onecolumngrid
\begin{center}
\begin{table}[h]
\setlength{\tabcolsep}{3mm}{
\begin{center}
\begin{tabular}{c|c|c|c|c}
  \hline\hline
 & $0.01\leq\eta\leq 0.02$ & $0.03\leq\eta\leq 0.36$ & $0.37\leq\eta\leq 0.88$ & $0.89\leq\eta\leq 1$\\
\hline
Existence of zeros of $\beta_3$ & \xmark &\cmark  & \cmark & \cmark\\
\hline
Existence of zeros of $\beta_4+16Q^2/(\pi T)$ & \xmark & \xmark  &   \cmark & \cmark  \\
\hline

Coexistence of zeros at the same temperature & \xmark &  \xmark & \xmark & \cmark
  \\
\hline

Intersection of zero loci in the $(m,n)$ plane & \xmark & \xmark  & \xmark & \xmark
  \\
  \hline\hline
\end{tabular}
\caption{Summary of the zero-point structure of $\beta_3$ and
$\beta_4+16Q^2/(\pi T)$ for different ranges of the occupancy
fraction $\eta$. A check mark indicates that the corresponding
condition is satisfied within the scanned parameter range, whereas
a cross indicates that it is not. Although zeros of both quantities
may coexist at the same temperature for $0.89\leq\eta\leq1$, their
zero loci do not intersect in the scanned $(m,n)$ parameter plane
at any fixed temperature.}\label{tab1}
\end{center}}
\end{table}
\end{center}
\twocolumngrid

\section{Conclusions and discussions}

In this work, we have examined whether the generalized \(m\)-\(n\) Lennard-Jones potential can provide a microscopic pair-interaction model that is simultaneously consistent with the Ruppeiner geometry and the equation of state of the charged AdS black hole. The equation of state was recast in terms of the virial expansion, so that the proposed interaction potential is constrained order by order through the corresponding Mayer cluster integrals. In contrast to the direct calculation of virial coefficients for a prescribed interaction potential, the present problem is an inverse one, in which the potential parameters are inferred from the required thermodynamic constraints. Directly solving for all four parameters \(m\), \(n\), \(d\), and \(E\) would therefore require solving a coupled set of nonlinear multidimensional integral equations. To make this inverse problem computationally tractable, we adopted a parameter-scanning strategy. Specifically, we first chose a trial pair $(m,n)$ satisfying $30\geq m>n>3$, where $n>3$ is required for the convergence of the second virial coefficient, while 30 is chosen as a practical upper bound for the numerical investigation. For each trial pair, the length scale \(d\) was determined from the Ruppeiner-geometric constraint, while the potential depth \(E\) was subsequently fixed by matching the second virial coefficient. The resulting parameter set \((m,n,d,E)\) was then used to calculate the third- and fourth-order virial coefficients and to test whether the remaining conditions could be satisfied simultaneously.

We systematically investigated the remaining third- and fourth-order constraints over the $(m,n)$ parameter space for different geometric prescriptions characterized by the occupancy fraction $\eta$. For the three representative choices, $\eta=\pi/6$, $\sqrt{2}\pi/6$, and $1$, no common zero of $\beta_3$ and $\beta_4+16Q^2/(\pi T)$ is found. The same conclusion is obtained for the additional occupancy fractions examined. We have also examined the dependence on the black-hole charge and temperature. To examine the charge dependence, we compare different values of \(Q\) at the same reduced temperature \(T/T_c\), so that the product \(QT\) remains fixed. For fixed \(\eta\), \(m\), and \(n\), the Ruppeiner constraint gives \(d^3\propto Q\). Moreover, the second-virial condition can be written as
\(\beta_2=d^3F_2(m,n,E/T)\). Since both \(d^3\) and the required value \(\beta_2=1/(\pi T)\) scale linearly with \(Q\), the ratio \(E/T\) remains independent of \(Q\). It then follows that $\beta_3\propto Q^2$ and $\beta_4\propto Q^3$. Since \(16Q^2/(\pi T)\) also scales as \(Q^3\) at fixed \(QT\), the quantity \(\beta_4+16Q^2/(\pi T)\) has the same \(Q^3\) scaling. Therefore, varying \(Q\) at fixed \(QT\) changes only the magnitudes of the relevant quantities, without altering their sign distributions or the locations of their zeros in the \((m,n)\) plane. The temperature scan further shows that the incompatibility is not specific to a particular temperature. For the three representative occupancy fractions, the zeros of 
\(\beta_3\) and \(\beta_4+16Q^2/(\pi T)\) either occur in different temperature ranges or, when they occur at the same temperature, their zero loci do not intersect in the \((m,n)\) parameter space. Extending the calculation to additional values of \(\eta\) between \(0\) and \(1\) likewise yields no common zero over the temperature range considered.

These results establish a virial-expansion no-go result for the generalized Lennard-Jones description considered here. Although such a potential can qualitatively represent the competition between short-range repulsion and longer-range attraction, it cannot simultaneously reproduce the complete virial structure of the charged AdS black-hole equation of state and satisfy the microscopic constraint inferred from the Ruppeiner geometry. Our analysis is based on the phenomenological interpretation that positive and negative Ruppeiner scalar curvatures are associated with repulsive and attractive microscopic interactions, respectively. This correspondence provides a qualitative characterization rather than an exact 
quantitative description of the underlying microscopic 
interactions. Therefore, the conclusions obtained in this 
work should also be understood within the framework of the 
adopted phenomenological assumptions.

The physical meaning of this incompatibility can be further understood from the roles of the individual virial coefficients. Matching the second and third virial coefficients, $\mathcal{V}_2$ and $\mathcal{V}_3$, ensures agreement with the charged AdS black-hole equation of state through order $v^{-3}$, thereby reproducing the correct low-density asymptotic behavior. Such a matching captures both the leading attractive correction associated with \(\mathcal{V}_2\) and the absence of a \(v^{-3}\) term. However, if \(\mathcal{V}_4\) is not reproduced, the model fails to capture the charge-induced repulsive contribution proportional to \(Q^2\). This interpretation is also consistent with the attractive microscopic interaction reported for the Schwarzschild-AdS (SAdS) black hole, for which the charge-dependent contribution is absent~\cite{AU,AV}. Since this fourth-order term competes with the attractive second-order contribution and is essential for generating the van der Waals-type inflection structure, failure to match \(\mathcal{V}_4\) prevents the model from reproducing the correct attraction-repulsion balance and the associated small/large black-hole phase transition.

Conversely, matching \(\mathcal{V}_2\) and \(\mathcal{V}_4\) preserves the leading attractive contribution and the charge-induced repulsive contribution responsible for the qualitative phase structure and critical behaviour. Nevertheless, a nonvanishing \(\mathcal{V}_3\) introduces a spurious \(v^{-3}\) term that is absent from the charged AdS black-hole equation of state. This additional contribution destroys the order-by-order low-density asymptotic agreement and generally shifts the critical volume, temperature, and pressure. Thus, matching \(\mathcal{V}_2\) and \(\mathcal{V}_3\) but not \(\mathcal{V}_4\) yields a more accurate low-density asymptotic description while missing the essential repulsive and critical contributions, whereas matching \(\mathcal{V}_2\) and \(\mathcal{V}_4\) but not \(\mathcal{V}_3\) retains the qualitative attraction-repulsion competition at the cost of introducing an unphysical intermediate-order term and modifying the quantitative critical structure. Only the simultaneous reproduction of \(\mathcal{V}_2\), \(\mathcal{V}_3\), and \(\mathcal{V}_4\) can recover both the complete virial structure and the correct critical behavior of the charged AdS black hole.

Within the framework of classical pairwise-additive interactions and the adopted prescriptions based on Ruppeiner thermodynamic geometry, our results demonstrate that generalized LJ potentials cannot consistently reproduce the virial coefficients required by the equation of state of charged AdS black holes over the investigated range of exponents. This conclusion reflects the limitations of the generalized LJ model under these assumptions, rather than excluding more general microscopic interaction models. In particular, other forms of microscopic interaction potentials beyond the generalized LJ model, non-pairwise-additive interactions, and alternative geometric prescriptions may provide different approaches to describing the microscopic interactions of charged AdS black holes. These results thus suggest a natural direction for future investigation. It would be interesting to explore more general microscopic interaction models and examine whether they can simultaneously reproduce the required virial coefficients and satisfy the constraint inferred from the Ruppeiner geometry. Such studies may provide further insight into the effective microscopic interactions underlying the thermodynamic behavior of charged AdS black holes and, more broadly, help establish a connection between the macroscopic equation of state, thermodynamic geometry, and the corresponding microscopic degrees of freedom.

{\it Acknowledgements}---During the preparation of this manuscript, C.H. Liu used ChatGPT to improve grammar and to assist with code writing.

%%%%%%%%%%%%%%%%%%%%%%%%%%%%%%%%%%%%%%%%%%%%%%%%%%%%%%%%%

%%%%%%%%%%%%%%%%%%%%%%%%%%%%%%%%%%%%%%%%%%%%%%%%%%%%%%%%%
%%%%%%%%%%%%%%%%%%%%%%%%%%%%%%%%%%%%%%%%%%%%%%%%%%%%%%%%%

%\newpage

%%%%%%%%%%%%%%%%%%%%%%%%%%%%%%%%%%%%%%%%%%%%%%%%%%%%%%%%%
%%%%%%%%%%%%%%%%%%%%%%%%%%%%%%%%%%%%%%%%%%%%%%%%%%%%%%%%%

%%%%%%%%%%%%%%%%%%%%%%% 
%%%%%%%%%%%%%%%%%%%%%%%%%%%%%%%%%%%%%%%%%%%%%%%%%%%%%%%%%
%%%%%%%%%%%%%%%%%%%%%%%%%%%%%%%%%%%%%%%%%%%%%%%%%%%%%%%%%


\begin{thebibliography}{99}
\bibitem{AA}
S. W. Hawking, \emph{Particle creation by black holes}, Commun. Math. Phys. {\bf 43} (3): 199 (1975).
%2
\bibitem{AB}
J. D. Bekenstein, \emph{Black holes and entropy}, Phys. Rev. {\bf D 7} (8): 2333 (1973).
%3
\bibitem{AC}
S. W. Hawking and D. N. Page, \emph{Thermodynamics of black holes in anti-de Sitter space}, Commun Math. Phys. {\bf 87} (4): 577 (1982).
%4
\bibitem{AD}
E. Witten, \emph{Anti-de Sitter space and holography}, Adv. Theor. Math. Phys. {\bf 2}: 253 (1998).
%5
\bibitem{AE}
A. Chamblin, R. Emparan, C. V. Johnson and R. C. Myers, \emph{Charged AdS black holes and catastrophic holography}, Phys. Rev. {\bf D 60} (6): 064018 (1999).
%6
\bibitem{AF}
A. Chamblin, R. Emparan, C. V. Johnson and R. C. Myers, \emph{Holography, thermodynamics and fluctuations of charged AdS black holes}, Phys. Rev. {\bf D 60} (10): 104026 (1999).
%7
\bibitem{AG}
D. Kastor, S. Ray and J. Traschen, \emph{Enthalpy and the Mechanics of AdS Black Holes}, Class. Quant. Grav. {\bf 26} (19): 195011 (2009).
%8
\bibitem{AH}
B. P. Dolan, \emph{The cosmological constant and the black hole equation of state}, Class. Quant. Grav. {\bf 28} (12): 125020 (2011).
%9
\bibitem{AI}
B. P. Dolan, \emph{Pressure and volume in the first law of black hole thermodynamics}, Class. Quant. Grav. {\bf 28} (23): 235017 (2011).
%10
\bibitem{AJ}
D. Kubiznak and R. B. Mann, \emph{P-V criticality of charged AdS black holes}, J. High Energy Phys. {\bf 2012} (7): 033 (2012).
%11
\bibitem{AK}
D. Kubiznak, R. B. Mann and M. Teo, \emph{Black hole chemistry: thermodynamics with Lambda}, Class. Quant. Grav. {\bf 34} (6): 063001 (2017).
%12
\bibitem{AL}
N. Altamirano, D. Kubiznak, R. B. Mann and Z. Sherkatghanad, \emph{Kerr-AdS analogue of triple point and solid/liquid/gas phase transition}, Class. Quant. Grav. {\bf 31} (4): 042001 (2014).
%13
\bibitem{AM}
R. A. Hennigar and R. B. Mann, \emph{Superfluids Black Holes}, Phys. Rev. Lett. {\bf 118} (2): 021301 (297).
%14
\bibitem{AN}
E. Spallucci and A. Smailagic, \emph{Maxwell's equal area law for charged Anti-de Sitter black holes}, Phys. Lett. {\bf B 723} (4): 436 (2013).


\bibitem{ZZZZA}
R. Li and J. Wang, \emph{Thermodynamics and kinetics of Hawking-Page phase transition}, Phys. Rev. D {\bf 102} (2): 024085 (2020).
%19
\bibitem{ZZZZB}
R. Li, K. Zhang, and J. Wang, \emph{Thermal dynamic phase transition of Reissner-Nordstr\"om Anti-de Sitter black holes on free energy landscape}, J. High Energy Phys. {\bf 2020} (10): 090 (2020).

\bibitem{ZZZZC}
S. J. Yang, R. Zhou, S. W. Wei, and Y. X. Liu, \emph{Kinetics of a phase transition for a Kerr-AdS black hole on the free-energy landscape}, Phys. Rev. D {\bf 105} (8): 084030 (2022).
%21
\bibitem{ZZZZD}
S. W. Wei, Y. Q. Wang, Y. X. Liu, and R. B. Mann, \emph{Observing dynamic oscillatory behavior of triple points among black hole thermodynamic phase transitions}, Sci. China Phys. Mech. Astron. {\bf 64} (7): 270411 (2021).
%22
\bibitem{ZZZZE}
J. X. Mo and S. Q. Lan, \emph{Dynamic phase transition of charged dilaton black holes}, Chin. Phys. C {\bf 45} (10): 105106 (2021).
%24
\bibitem{ZZZZF}
R. Li, K. Zhang, and J. Wang, \emph{Kinetics and its turnover of Hawking-Page phase transition under the black hole evaporation}, Phys. Rev. D {\bf 104} (8): 084060 (2021).
%25
\bibitem{ZZZZG}
R. Li, K. Zhang, and J. Wang, \emph{Probing black hole microstructure with the kinetic turnover of phase transition}, Phys. Rev. D {\bf 104} (8): 084076 (2021).
%26
\bibitem{ZZZZH}
R. Li and J. Wang, \emph{Kinetics of Hawking-Page phase transition with the non-Markovian effects}, J. High Energy Phys. {\bf 2022} (5): 128 (2022).
%30
\bibitem{ZZZZI}
S. W. Wei, Y. X. Liu, and Y. Q. Wang, \emph{Dynamic properties of thermodynamic phase transition for five-dimensional neutral Gauss-Bonnet AdS black hole on free energy landscape}, Nucl. Phys. B {\bf 976}: 115692 (2022).
%31
\bibitem{ZZZZJ}
J. Y. Yang and R. B. Mann, \emph{Dynamic behaviours of black hole phase transitions near quadruple points}, J. High Energy Phys. {\bf 2023} (8): 028 (2023).
\bibitem{ZZZZK}
Z. M. Xu, \emph{Fokker-Planck equation for black holes in thermal potential},  Phys. Rev. D {\bf 104} (10) : 104022 (2021).
\bibitem{ZZZZL}
C. Liu and J. Wang, \emph{Path integral and instantons for the dynamical process and phase transition rate of Reissner-Nordstr\"om-AdS black holes}, Phys. Rev. D {\bf 105} (10): 104024 (2022).
\bibitem{ZZZZM}
C. Liu, R. Li, K. Zhang, and J. Wang, \emph{Generalized free energy and dynamical state transition of the dyonic AdS black hole in the grand canonical ensemble}, J. High Energy Phys. {\bf 2023} (11): 068 (2023).
\bibitem{ZZZZN}
Z. M. Xu, B. Wu, and W. L. Yang, \emph{Rate of the phase transition for a charged anti-de Sitter black hole}, Sci. China Phys. Mech. Astron. {\bf 66} (4): 240411 (2023).
\bibitem{ZZZZO}
Y. S. Wang, Z. M. Xu, and B. Wu, \emph{Thermodynamic phase transition rate for the third-order Lovelock black hole in diverse dimensions}, Phys. Lett. B {\bf 853}: 138690 (2024).
\bibitem{ZZZZP}
C. Ma, P. P. Zhang, B. Wu, and Z. M. Xu, \emph{The Kramers escape rate of phase transitions for the 6-dimensional Gauss-Bonnet AdS black hole with triple phases}, Phys. Lett. B {\bf 861}: 139282 (2025).
\bibitem{ZZZZQ}
C. Ma, B. Wu, and Z. M. Xu, \emph{Mean first passage time and the Kramers escape rate of phase transitions for the Bardeen-AdS-class black hole}, Chin. Phys. C {\bf 50} (8): 085103 (2026).
\bibitem{ZZZZR}
R. Li and J. Wang, \emph{Generalized free energy landscape of black hole phase transition}, Phys. Rev. D {\bf 106} (10): 106015 (2022).
\bibitem{ZZZZS}
S. P. Wu, Y. X. Liu, and S. W. Wei, \emph{Generalized free energy landscapes from Iyer-Wald formalism}, Phys. Dark Univ. {\bf 51}: 102210 (2026).

\bibitem{ZZZA}
S. W. Wei, Y. X. Liu, and R. B. Mann, \emph{Black Hole Solutions as Topological Thermodynamic Defects}, Phys. Rev. Lett. {\bf 129} (19): 191101 (2022).
\bibitem{ZZZB}
P. K. Yerra, C. Bhamidipati, and S. Mukherji, \emph{Topology of critical points and Hawking-Page transition}, Phys. Rev. D {\bf 106} (6): 064059 (2022).
%35
\bibitem{ZZZC}
C. H. Liu and J. Wang, \emph{Topological natures of the Gauss-Bonnet black hole in AdS space}, Phys. Rev. D {\bf 107} (6): 064023 (2023).
%36
\bibitem{ZZZD}
R. Li, C. H. Liu, K. Zhang, and J. Wang, \emph{Topology of the landscape and dominant kinetic path for the thermodynamic phase transition of the charged Gauss-Bonnet-AdS black holes}, Phys. Rev. D {\bf 108} (4): 044003 (2023).
%37
\bibitem{ZZZE}
C. X. Fang, J. Jiang, and M. Zhang, \emph{Revisiting thermodynamic topologies of black holes}, J. High Energy Phys. {\bf 2023} (1): 102 (2023).
%36
\bibitem{ZZZF}
D. Wu, \emph{Topological classes of rotating black holes}, Phys. Rev. D {\bf 107} (2): 024024 (2023).
%41
\bibitem{ZZZG}
Y. B. Du and X. D. Zhang, \emph{Topological classes of black holes in de-Sitter spacetime}, Eur. Phys. J. C {\bf 83} (10): 927 (2023).
%47
\bibitem{ZZZH}
M. S. Ali, H. El Moumni, J. Khalloufi, and K. Masmar, \emph{Topology of Born-Infeld-AdS black hole phase transitions: Bulk and CFT sides}, Annals Phys. {\bf 465}: 169679 (2024).
%50
\bibitem{ZZZI}
D. Y. Chen, Y. C. He, and J. Tao, \emph{Topological classes of higher-dimensional black holes in massive gravity}, Eur. Phys. J. C {\bf 83} (9): 872 (2023).
%51
\bibitem{ZZZJ}
J. Sadeghi, M. R. Alipour, S. N. Gashti, and M. A. S. Afshar, \emph{Bulk-boundary and RPS thermodynamics from topology perspective}, Chin. Phys. C {\bf 48} (9): 095106 (2024).
\bibitem{ZZZK}
S. W. Wei and Y. X. Liu, \emph{Topology of black hole thermodynamics: A brief review}, Sci. China Phys. Mech. Astron. {\bf 69} (6): 260401 (2026).

\bibitem{ZZZL}
S. J. Yang, S. P. Wu, S. W. Wei, and Y. X. Liu, \emph{Topological perspective on bulk boundary thermodynamic equivalence}, Eur. Phys. J. C {\bf 86} (5): 498 (2026).

\bibitem{ZZZM}
S. W. Wei, Y. X. Liu, and R. B. Mann, \emph{Universal topological classifications of black hole thermodynamics}, Phys. Rev. D {\bf 110} (8): L081501 (2024).

\bibitem{ZZZN}
S. P. Wu, S. J. Yang, and S. W. Wei, \emph{Extended thermodynamical topology of black hole}, Eur. Phys. J. C {\bf 85} (12): 1372 (2025).

\bibitem{ZZZO}
Y. S. Wang, Z. M. Xu, and B. Wu, \emph{Thermodynamic phase transition and winding number for the third-order Lovelock black hole}, Chin. Phys. C {\bf 48} (9): 095101 (2024).

\bibitem{ZZZP}
D. Wu, W. Liu, S. Q. Wu, and R. B. Mann, \emph{Novel topological classes in black hole thermodynamics}, Phys. Rev. D {\bf 111} (6): L061501 (2025).
\bibitem{ZKA}
Z. M. Xu and R. B. Mann, \emph{Thermodynamic Supercriticality and Complex Phase Diagram for the AdS Black Hole}, Phys. Rev. Lett. {\bf 136} (4): 041402 (2026).
\bibitem{ZKB}
Z. Y. Li, X. R. Chen, B. Wu, and Z. M. Xu, \emph{Thermodynamic supercriticality and complex phase diagram for charged Gauss-Bonnet AdS black holes}, Phys. Lett. B {\bf 876}: 140438 (2026).

\bibitem{ZKC}
F. Guo and Z. M. Xu, \emph{Complex plane phase diagram and Widom line for the Born-Infeld black holes with reentrant phase transition}, Eur. Phys. J. C {\bf 86} (5): 524 (2026).

\bibitem{ZKD}
J. Y. Xia and Z. M. Xu, \emph{Thermodynamic Supercritical Scaling Law for the charged Gauss-Bonnet AdS Black Hole}, arXiv:2610.27824 (2026).

\bibitem{ZKE}
D. Wu and X. D. Zhu, \emph{Complex phase diagram and thermodynamic supercriticality of five-dimensional four-charge AdS black holes in $STU-W^2U$ gauged supergravity}, arXiv:2611.05624 (2026).


\bibitem{ZKF}
Z. Q. Zhao, Z. Y. Nie, J. F. Zhang, and X. Zhang, \emph{Characterized behaviors of black hole thermodynamics in the supercritical region}, Chin. Phys. Lett. {\bf 42} (10): 101102 (2025).
\bibitem{ZKG}
R. Li, K. Zhang, J. Y. Yang, R. B. Mann, and J. Wang, \emph{Critical slowing down of black hole phase transition and kinetic crossover in supercritical regime}, Phys. Rev. D {\bf 112} (6): 064004 (2025).

\bibitem{ZKH}
S. C. Wang, X. Y. Li, Y. L. Jin, and L. Li, \emph{Universal Supercritical Thermodynamics for Black Holes}, Research {\bf 9}: 1406 (2026).

\bibitem{ZKI}
A. Anand and S. C. Wang, \emph{Universal supercritical behavior in global monopole-charged AdS black holes}, J. High Energy Phys. {\bf 2026} (5): 278 (2026).

%15
\bibitem{AO}
S.-W. Wei and Y.-X. Liu, \emph{Insight into the microscopic structure of an AdS black hole from thermodynamical phase transition}, Phys. Rev. Lett. {\bf 115} (11): 111302 (2015).
%16
\bibitem{AP}
S.-W. Wei, Y.-X. Liu and R. B. Mann, \emph{Repulsive Interactions and University properties of Charged AdS Black Hole Microstructures}, Phys. Rev. Lett. {\bf 123} (7): 071103 (2019).
%17
\bibitem{AQ}
G. Ruppeiner, \emph{Riemannian geometry in thermodynamic fluctuation theory}, Rev. Mod. Phys. {\bf 67} (3): 605 (1995).
%18
\bibitem{AR}
G. Ruppeiner, \emph{Thermodynamics: A Riemannian geometric model}, Phys. Rev. {\bf A 20} (4): 1608 (1979).
%19
\bibitem{AS}
G. Ruppeiner, \emph{Thermodynamic curvature measures interactions}, Am. J. Phy, {\bf 78} (11): 1170 (2010).
%20
\bibitem{AT}
S.-W. Wei, Y.-X. Liu and R. B. Mann, \emph{Ruppeiner Geometry, Phase Transitions, and the Microstructure of Charged AdS Black Holes}, Phys. Rev. {\bf D 100} (12): 124033 (2019).
%21
\bibitem{AU}
S.-W. Wei, Y.-X. Liu and R. B. Mann, \emph{Novel dual relation and constant in Hawking-Page phase transitions}, Phys. Rev. {\bf D 102} (10): 104011 (2020).
%22
\bibitem{AV}
Z.-M. Xu, B. Wu and W.-L. Yang, \emph{Ruppeiner thermodynamic geometry for the Schwarzschild-AdS black hole}, Phys. Rev. {\bf D 101} (2): 024018 (2020).
%23
\bibitem{AW}
Y.-G. Miao and Z.-M. Xu, \emph{Thermal molecular potential among micromolecules in charged AdS black holes}, Phys. Rev. {\bf D 98} (4): 044001 (2018).
%24
\bibitem{AX}
S.-W. Wei, Y.-X. Liu and R. B. Mann, \emph{Characteristic interaction potential of black hole molecules from the microscopic interpretation of Ruppeiner geometry}, [arxiv: 2108. 07655 [gr-qc]].
%29
\bibitem{BD}
C. H. Liu and J. Wang, \emph{The radial distribution function reveals the underlying mesostructure of the AdS black hole}, J. High Energy Phys. {\bf 2022} (10): 171 (2012).
\bibitem{KKKA}
Y. G. Miao and Z. M. Xu, \emph{Interaction potential and thermo-correction to the equation of state for thermally stable Schwarzschild Anti-de Sitter black holes}, Sci. China Phys. Mech. Astron. {\bf 62} (1): 10412 (2019).
%26
\bibitem{BA}
H. D. Ursell, \emph{The evaluation of Gibbs' phase-integral for imperfect gases}, Proc. Camb. Phil. Soc. 23 (6): 685 (1927).
%28
\bibitem{BC}
J. A. Barker and J. J. Monaghan, \emph{Fourth Virial Coefficients for the $12-6$ Potential}, J. Chem. Phys. {\bf 36} (10): 2564 (1962).

%
\bibitem{CC}
 See Supplemental Material for the details.






\end{thebibliography}
\end{document}

\begin{CJK*}{UTF8}{gbsn}

\title{Supplemental Materials for ``Failure of Generalized Lennard-Jones Models for Charged AdS Black Hole Molecules in the Virial Expansion"}

\author{Conghua Liu (刘聪华)}
\affiliation{College of Physics, Jilin University, Changchun 130012, China}

\author{Jin Wang (汪劲)}
\email{jin.wang.1@stonybrook.edu}
\affiliation{Department of Chemistry and Department of Physics and Astronomy, State University of New York at Stony Brook, Stony Brook, New York 11794, USA}

\maketitle
\end{CJK*}
\onecolumngrid

\tableofcontents

%%%%%%%%%%%%%%%%%%%%%%%%%%%%%%%%%%%%%%%%%%%%%%%%%%%%%%%%%

\onecolumngrid

\section{Derivation of the Virial-Coefficient Integrals}
We set the particle 0 as the original point, and $r_{01}$ as $r_1$, $r_{02}$ as $r_2$.
\be
\ba
\beta_2&=\int f_{01} d\mathbf{r}_1\\
&=\int_0^{\infty} {r_1}^2f(r_1)  dr_1\int_0^{\pi}\sin\theta d\theta \int_0^{2\pi}d\phi\\
&=\int_0^{\infty} 4\pi {r_1}^2 f(r_1) dr_1.
\ea
\ee
Regarding $\beta_3$, we write
\be
\ba
\beta_3=\int\int f_{01}f_{02}f_{12}d\mathbf{r}_1d\mathbf{r}_2
\ea
\ee
Note that $r_{01}=r_1$, $r_{02}=r_2$, the angle between $\mathbf{r}_1$ and $\mathbf{r}_2$ is $\theta_{12}$, we can write
\be
\ba
&\beta_3=\int\int f_{01}f_{02}f_{12}d\mathbf{r}_1d\mathbf{r}_2\\
&=\int f(r_1) d\mathbf{r}_1 \int f(r_2)f(r_{12})d\mathbf{r}_2\\
&=\int_0^{\infty} dr_1 4\pi r_1^2 f(r_1) \int_0^{\infty}dr_2 2\pi r_2^2 f(r_2)\int_0^{\pi}d\theta_{12} f(r_{12}) \sin\theta_{12}.
\ea
\ee
The expression can be simplified if one introduces a function $\kappa(r_1)=\int f(r_2)f(r_{12})d\mathbf{r}_2$, which can be interpreted as the convolution of the Mayer function with itself.
From a physical perspective, it quantifies the indirect correlation between two particles arising from their interactions with a third particle. One can choose $\mathbf{r_1}$ as the $z$-axis, and the result can be simplified as:
\be
\ba
\kappa(r_1)=2\pi\int_0^{\infty} f(r_2){r_2}^2dr_2 \int_0^{\pi}f(r_{12})\sin\theta_{12}d\theta_{12},
\ea
\ee
Then, 
\be
\ba
\beta_3=4\pi\int_0^{\infty} dr_1 r_1^2 f(r_1)\kappa(r_1).
\ea
\ee
Then, we simplify the expression of $\kappa(r_1)$. The cosine law shows that
\be
\ba
\label{eq:Ici4}
r_{12}^2={r_1}^2+{r_2}^2-2r_1r_2\cos\theta_{12},
\ea
\ee
we have $r_{12}dr_{12}=r_1r_2\sin\theta_{12}d\theta_{12}$. Thus,
\be
\ba
&\kappa(r_1)=\\
&2\pi\int_0^{\infty} f(r_2){r_2}^2dr_2 \int_{|r_1-r_2|}^{|r_1+r_2|}f(r_{12})\sin\theta_{12}\frac{r_{12}}{r_1r_2\sin\theta_{12}}dr_{12}\\
&=2\pi\int_0^{\infty} f(r_2){r_2}dr_2 \int_{|r_1-r_2|}^{|r_1+r_2|}f(r_{12})\frac{r_{12}}{r_1}dr_{12}.
\ea
\ee
Then, we have 
\be
\ba
\beta_3=8\pi^2\int_0^{\infty} dr_1 r_1 f(r_1)\int_0^{\infty} f(r_2){r_2}dr_2 \int_{|r_1-r_2|}^{|r_1+r_2|}f(r_{12}){r_{12}}dr_{12}
\ea
\ee

One can then choose $\mathbf{r_2}$ as the $z$-axis and analyze $\beta_4^{(1)}$, which is written as
\be
\ba
\beta_4^{(1)}&=\int\int\int f_{01}f_{12}f_{23}f_{30}d\mathbf{r}_1d\mathbf{r}_2d\mathbf{r}_3\\
&=\int d\mathbf{r}_2 \int d\mathbf{r}_1 f(r_1)f(r_{21})\int d\mathbf{r}_3 f(r_3)f(r_{32})\\
&=4\pi\int_0^{\infty} dr_2 r_2^2[\kappa(r_2)]^2\\
&=16\pi^3\int_0^{\infty} dr_2 \big[\int_0^{\infty} f(r_1){r_1}dr_1 \int_{|r_1-r_2|}^{|r_1+r_2|}f(r_{12}){r_{12}}dr_{12}\big]^2.
\ea
\ee
In addition, $\beta_4^{(2)}$ can be written as
\be
\ba
\beta_4^{(2)}&=\int\int\int f_{01}f_{12}f_{23}f_{30}f_{02}d\mathbf{r}_1d\mathbf{r}_2d\mathbf{r}_3\\
&=\int d\mathbf{r}_2 f(r_2) \int d\mathbf{r}_1 f(r_1)f(r_{21})\int d\mathbf{r}_3 f(r_3)f(r_{32})\\
&=4\pi\int_0^{\infty} dr_2 f(r_2)r_2^2[\kappa(r_2)]^2\\
&=16\pi^3\int_0^{\infty} dr_2 f(r_2)[\int_0^{\infty} f(r_1){r_1}dr_1 \int_{|r_1-r_2|}^{|r_1+r_2|}f(r_{12}){r_{12}}dr_{12}]^2,
\ea
\ee
where 
\be
\ba
\kappa(r_2)&=2\pi\int_0^{\infty} f(r_1){r_1}^2dr_1 \int_0^{\pi}f(r_{21})\sin\theta_{21}d\theta_{21}\\
&=2\pi\int_0^{\infty} f(r_1){r_1}dr_1 \int_{|r_1-r_2|}^{|r_1+r_2|}f(r_{12})\frac{r_{12}}{r_2}dr_{12}.
\ea
\ee

Finally, we analyze $\beta_4^{(3)}$, which is written as
\be
\ba
\beta_4^{(3)}=&\int\int\int f_{01}f_{12}f_{23}f_{30}f_{13}f_{02}d\mathbf{r}_1d\mathbf{r}_2d\mathbf{r}_3,
\ea
\ee
where $r_{13}=r_1^2+r_3^2-2r_1r_3\cos\theta_{13}$. We denote the angle between the planes $012$ and $023$ as $\phi$, which satisfies:
\be
\ba
\label{eq:Ici7}
\cos\theta_{13}=\cos\theta_{12}\cos\theta_{23}+\sin\theta_{12}\sin\theta_{23}\cos\phi,
\ea
\ee
thus $\theta_{13}$ depends on $\phi$, and we can not directly integral $\phi$ as $2\pi$. Then, we can rewrite $\beta_4^{(3)}$ as
\be
\ba
\label{dnjsndj}
\beta_4^{(3)}=&8\pi^2\int_0^{\infty}f_{01}r_{01}^2 dr_{01}\int_0^{\infty}f_{02}r_{02}^2dr_{02}\int_0^{\infty}f_{30}r_{03}^2dr_{03}\\
&\int_0^{\pi} d\theta_{12}\sin\theta_{12}\int_0^{\pi} d\theta_{23}\sin\theta_{23}\int_0^{2\pi} d\phi f_{12}f_{23}f_{13}
\ea
\ee
Here, $f_{12}$, $f_{23}$, and $f_{13}$ can be expanded in terms of Legendre polynomials as:
\be
\ba
\label{jdishk}
f_{12}&=\sum_{l=0}^{\infty} A_l(r_1, r_2)P_l(\cos\theta_{12}),\\
f_{23}&=\sum_{m=0}^{\infty} A_m(r_2, r_3) P_m(\cos\theta_{23}),\\
f_{13}&=\sum_{n=0}^{\infty} A_n(r_1, r_3) P_n(\cos\theta_{13}),
\ea
\ee
where $\frac{2A_b(r_{0i},r_{0j})}{2b+1}=\int_{-1}^1 f_{ij} P_b(\cos\theta_{ij})d(\cos\theta_{ij})$. Using the addition theorem, $P_n(\cos\theta_{13})$ can be expressed by $P_l(\cos\theta_{12})$ and $P_m(\cos\theta_{23})$, written as:
\be
\ba
\label{dsffvdfc}
P_n(\cos\theta_{13})=P_n(\cos\theta_{12})P_n(\cos\theta_{23})+2\sum_{s=1}^n\frac{(n-s)!}{(n+s)!}P_n^s(\cos\theta_{12})P_n^s(\cos\theta_{23})\cos(s\phi).
\ea
\ee

Substituting Eq.~(\ref{jdishk}) into Eq.~(\ref{dnjsndj}), we obtain
\be
\ba
\beta_4^{(3)}=&8\pi^2\int_0^{\infty}f_{01}r_{01}^2 dr_{01}\int_0^{\infty}f_{02}r_{02}^2dr_{02}\int_0^{\infty}f_{30}r_{03}^2dr_{03}\int_0^{\pi} d\theta_{12}\sin\theta_{12}\int_0^{\pi} d\theta_{23}\sin\theta_{23}\int_0^{2\pi} d\phi\\
&[\sum_l A_l(r_1, r_2)P_l(\cos\theta_{12})][\sum_m A_m(r_2, r_3) P_m(\cos\theta_{23})][\sum_n A_n(r_1, r_3) P_n(\cos\theta_{13})].
\ea
\ee
Note that $\int_0^{2\pi}d\phi \cos(s\phi)=0$ for all $s=1,2,3,...$. Thus, we obtain
\be
\ba
\beta_4^{(3)}=&16\pi^3\int_0^{\infty}f_{01}r_{01}^2 dr_{01}\int_0^{\infty}f_{02}r_{02}^2dr_{02}\int_0^{\infty}f_{30}r_{03}^2dr_{03}\int_0^{\pi} d\theta_{12}\sin\theta_{12}\int_0^{\pi} d\theta_{23}\sin\theta_{23}\\
&[\sum_{l=0}^{\infty} A_l(r_1, r_2)P_l(\cos\theta_{12})][\sum_{m=0}^{\infty} A_m(r_2, r_3) P_m(\cos\theta_{23})][\sum_{n=0}^{\infty} A_n(r_1, r_3) P_n(\cos\theta_{12})P_n(\cos\theta_{23})]\\
=&16\pi^3\sum_{l=0}^{\infty} \sum_{m=0}^{\infty} \sum_{n=0}^{\infty}  \int_0^{\infty}f_{01}r_{01}^2 dr_{01}\int_0^{\infty}f_{02}r_{02}^2dr_{02}\int_0^{\infty}f_{30}r_{03}^2dr_{03} A_l(r_1, r_2) A_m(r_2, r_3) A_n(r_1, r_3)\\
&\int_0^{\pi} d\theta_{12}\sin\theta_{12}P_l(\cos\theta_{12})P_n(\cos\theta_{12})\int_0^{\pi} d\theta_{23}\sin\theta_{23}P_m(\cos\theta_{23})P_n(\cos\theta_{23}).
\ea
\ee
Note that 
\be
\ba
&\int_0^{\pi} d\theta_{12}\sin\theta_{12}P_l(\cos\theta_{12})P_n(\cos\theta_{12})=\int_{-1}^{1}d(\cos\theta_{12}) P_l(\cos\theta_{12})P_n(\cos\theta_{12})=\frac{2}{2n+1}\delta_{ln},\\
&\int_0^{\pi} d\theta_{23}\sin\theta_{23}P_m(\cos\theta_{23})P_n(\cos\theta_{23})=\int_{-1}^{1}d(\cos\theta_{23}) P_m(\cos\theta_{23})P_n(\cos\theta_{23})=\frac{2}{2n+1}\delta_{mn}.
\ea
\ee
One can obtain
\be
\ba
\beta_4^{(3)}= \sum_{n=0}^{\infty}\frac{64\pi^3}{(2n+1)^2} \int_0^{\infty}f_{01}r_{01}^2 dr_{01}\int_0^{\infty}f_{02}r_{02}^2dr_{02}\int_0^{\infty}f_{30}r_{03}^2dr_{03} A_n(r_1, r_2) A_n(r_2, r_3) A_n(r_1, r_3).
\ea
\ee

Here,
\be
\ba
\frac{2A_b(r_{i},r_{j})}{2b+1}=&\int_{-1}^1 f_{ij} P_b(\cos\theta_{ij})d(\cos\theta_{ij})\\
=& \frac{1}{r_ir_j}\int_{|r_i-r_j|}^{|r_i+r_j|}P_b(\frac{r_i^2+r_j^2-r_{ij}^2}{2r_ir_j})f(r_{ij})r_{ij}dr_{ij},
\ea
\ee
where we have used the relation $2r_{ij}dr_{ij}=-2r_ir_jd(\cos\theta_{ij})$ from $r_{ij}^2=r_i^2+r_j^2-2r_ir_j\cos\theta_{ij}$. Thus, $\beta_4^{(3)}$ is written as
\be
\ba
\beta_4^{(3)}=&\sum_{n=0}^{\infty}{8\pi^3(2n+1)} \int_0^{\infty}f(r_{1}) dr_{1}\int_0^{\infty}f(r_{2})dr_{2}\int_0^{\infty}f(r_{3})dr_{3}\int_{|r_1-r_2|}^{|r_1+r_2|}P_n(\frac{r_1^2+r_2^2-r_{12}^2}{2r_1r_2})f(r_{12})r_{12}dr_{12}\\
&\int_{|r_2-r_3|}^{|r_2+r_3|}P_n(\frac{r_2^2+r_3^2-r_{23}^2}{2r_2r_3})f(r_{23})r_{23}dr_{23}\int_{|r_1-r_3|}^{|r_1+r_3|}P_n(\frac{r_1^2+r_3^2-r_{13}^2}{2r_1r_3})f(r_{13})r_{13}dr_{13}.
\ea
\ee
Here, $P_0(x)=1$, $P_1(x)=x$, $P_2(x)=\frac{1}{2}(3x^2-1)$, $P_3(x)=\frac{1}{2}(5x^3-3x)$, $P_4(x)=\frac{1}{8}(35x^4-30x^2+3)$, and $P_5(x)=\frac{1}{8}(63x^5-70x^3+15x)$.

%%%%

\section{Additional numerical results}
Here, we provide additional numerical results verifying that, at general occupancy fractions, no common zero of $\beta_3$ and $\beta_4+16Q^2/(\pi T)$ is found at fixed temperature. These zeros fall into four cases. For $0.01 \leq \eta \leq 0.02$, neither $\beta_3$ nor $\beta_4+16Q^2/(\pi T)$ has any zeros. For $0.03 \leq \eta \leq 0.36$, the zero points of $\beta_3$ can occur, whereas no zeros of $\beta_4+16Q^2/(\pi T)$ can occur. For $0.37 \leq \eta \leq 0.88$, zeros of both $\beta_3$ and $\beta_4+16Q^2/(\pi T)$ can occur, but they do not occur at the same temperature. For $0.89 \leq \eta \leq 1$, zeros of both $\beta_3$ and $\beta_4+16Q^2/(\pi T)$ can occur, and they can occur at the same temperature. However, their zero loci in the $(m,n)$ plane do not intersect at any fixed temperature.

\begin{figure}[htp]
\includegraphics[width=16cm]{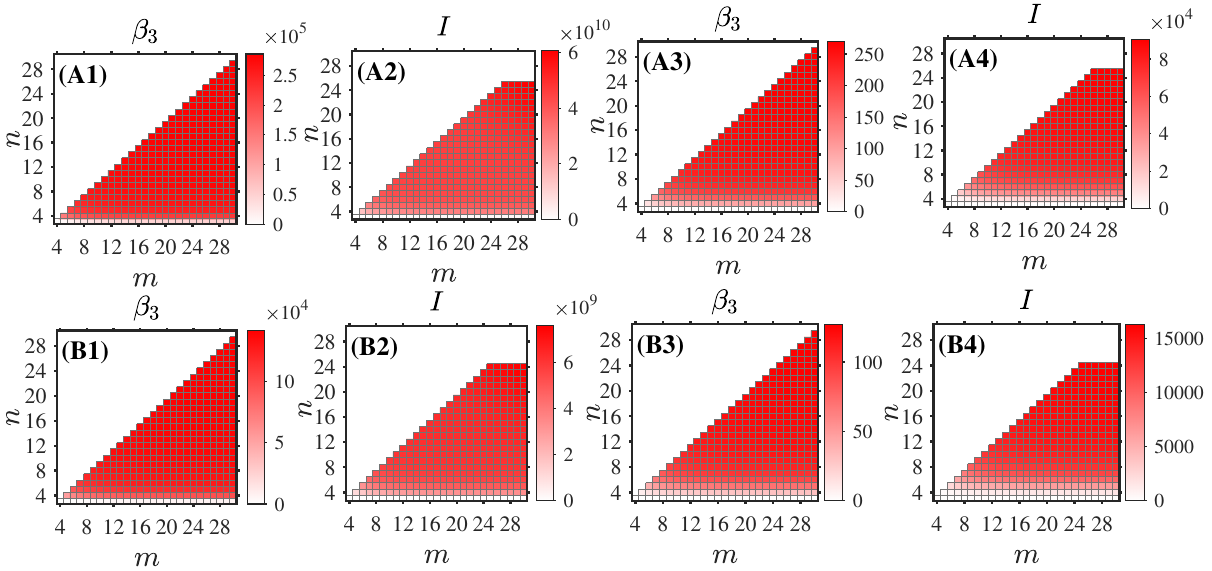}
\caption{Characteristic behaviors in the regime $0.01 \leq \eta \leq 0.02$ are shown. Here, we denote $\beta_4+16Q^2/(\pi T)$ by $I$. In this regime, neither $\beta_3$ nor $I$ has any zeros. Panels (A1)--(A4) correspond to $\eta=0.01$; panels (A1) and (A2) correspond to $T=0.1T_c$, while panels (A3) and (A4) correspond to $T=1.0T_c$. Panels (B1)--(B4) correspond to $\eta=0.02$; panels (B1) and (B2) correspond to $T=0.1T_c$, while panels (B3) and (B4) correspond to $T=1.0T_c$.}
\label{Fig10}
\end{figure}
%%%%%%%%%%%%%%%%%%%%%%%%%%%%%%%%%%%%%%%%%%%%%%%%%%%%%%%%%

\begin{figure}[t]
\includegraphics[width=16cm]{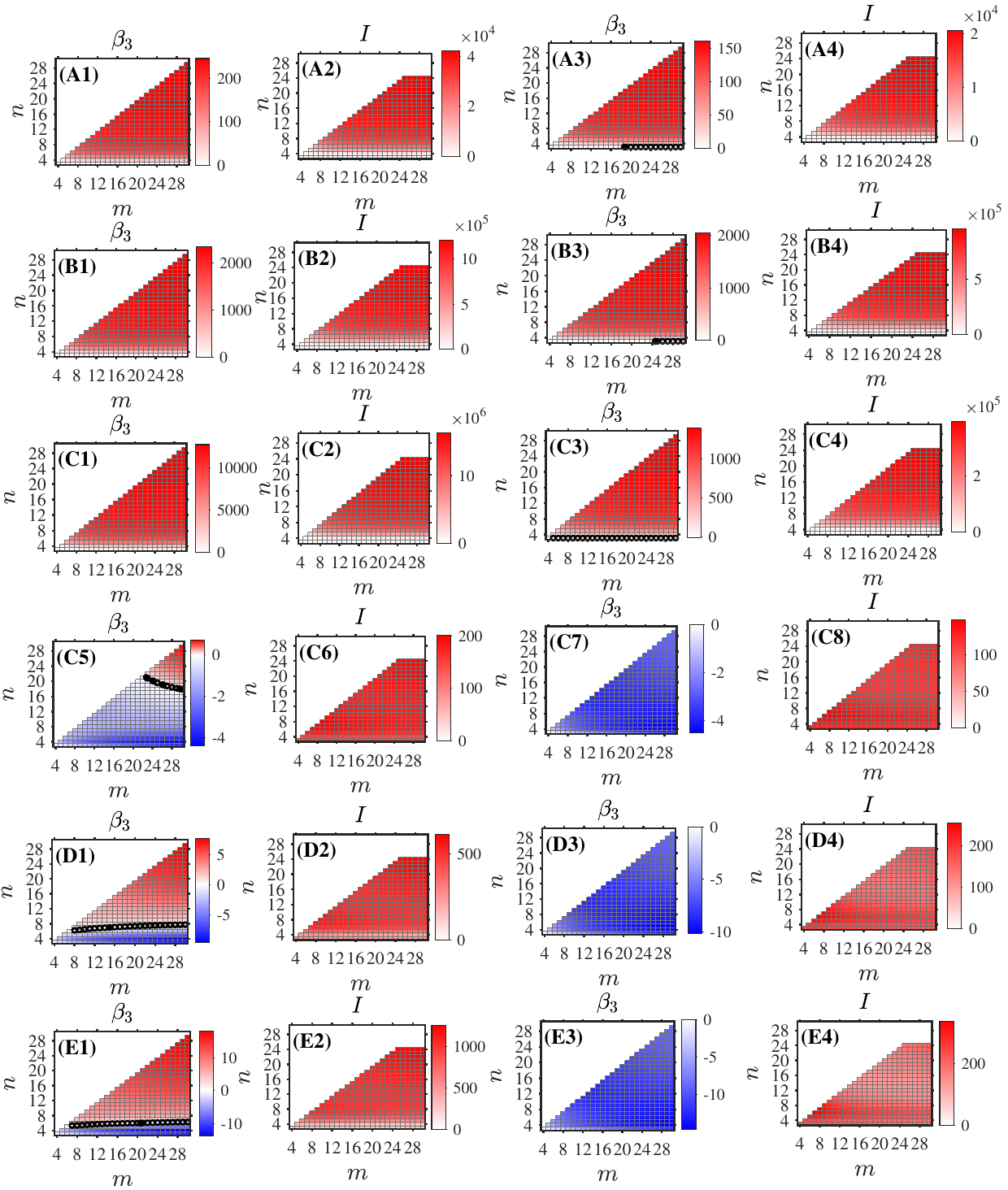}
\caption{Characteristic behaviors in the regime $0.03 \leq \eta \leq 0.36$ are shown. In this regime, the zero of $\beta_3$ can occur, whereas the zeros of $\beta_4+16Q^2/(\pi T)$ cannot occur. Panels (A1)--(A4) correspond to $\eta=0.03$. Panels (A1) and (A2) correspond to $T=0.7T_c$; at this $\eta$, the zero of $\beta_3$ is absent below this temperature. Panels (A3) and (A4) correspond to $T=0.8T_c$; at this $\eta$, the zero of $\beta_3$ always occurs above this temperature. Panels (B1)--(B4) correspond to $\eta=0.1$. Panels (B1) and (B2) correspond to $T=0.22T_c$; at this $\eta$, the zero of $\beta_3$ is absent below this temperature. Panels (B3) and (B4) correspond to $T=0.23T_c$; at this $\eta$, the zero of $\beta_3$ always occurs above this temperature. Panels (C1)--(C8) correspond to $\eta=0.2$. Panels (C1) and (C2) correspond to $T=0.1T_c$; at this $\eta$, the zero of $\beta_3$ is absent below this temperature. Panels (C3) and (C4) correspond to $T=0.2T_c$; at this $\eta$, the zero of $\beta_3$ always occurs above this temperature but below $0.8T_c$. Panels (C5) and (C6) correspond to $T=0.8T_c$, and panels (C7) and (C8) correspond to $T=0.9T_c$; at this $\eta$, the zero of $\beta_3$ is absent above $0.9T_c$. Panels (D1)--(D4) correspond to $\eta=0.3$. Panels (D1) and (D2) correspond to $T=0.5T_c$; at this $\eta$, the zero of $\beta_3$ always occurs below this temperature. Panels (D3) and (D4) correspond to $T=0.6T_c$; at this $\eta$, the zero of $\beta_3$ is absent above this temperature. Panels (E1)--(E4) correspond to $\eta=0.36$. Panels (E1) and (E2) correspond to $T=0.4T_c$; at this $\eta$, the zero of $\beta_3$ always occurs below this temperature. Panels (E3) and (E4) correspond to $T=0.5T_c$; at this $\eta$, the zero of $\beta_3$ is absent above this temperature.}
\label{Fig11}
\end{figure}
%%%%%%%%%%%%%%%%%%%%%%%%%%%%%%%%%%%%%%%%%%%%%%%%%%%%%%%%%

\begin{figure}[t]
\includegraphics[width=16cm]{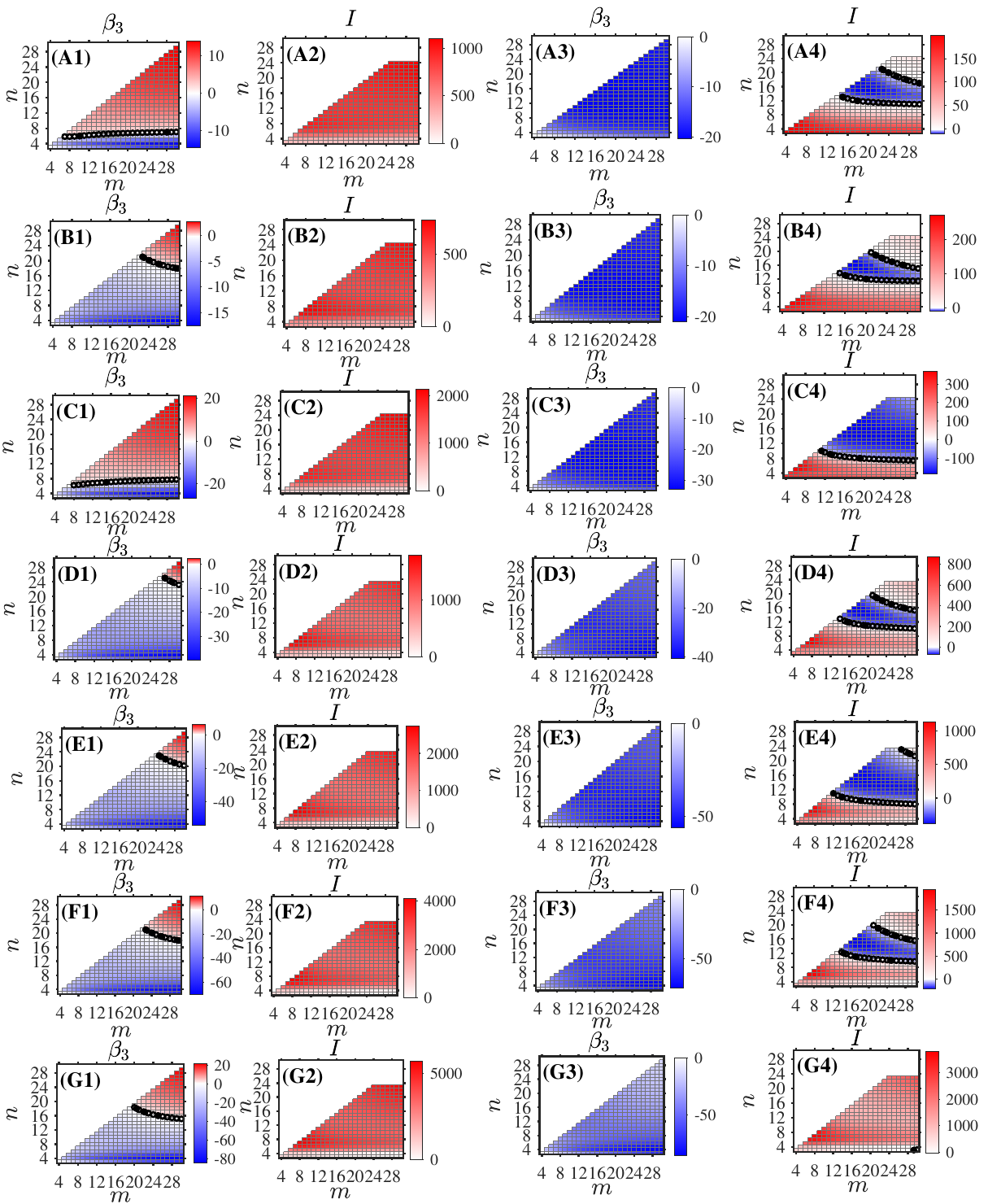}
\caption{Characteristic behaviors in the regime $0.37 \leq \eta \leq 0.88$ are shown. In this regime, the zeros of both $\beta_3$ and $\beta_4+16Q^2/(\pi T)$ can occur, but they do not occur at the same temperature. Panels (A1)--(A4) correspond to $\eta=0.37$. Panels (A1) and (A2) correspond to $T=0.4T_c$; at this $\eta$, the zero of $\beta_3$ always occurs below this temperature. Panels (A3) and (A4) correspond to $T=0.6T_c$; at this $\eta$, the zero of $\beta_4+16Q^2/(\pi T)$ always occurs above this temperature. Panels (B1)--(B4) correspond to $\eta=0.4$. Panels (B1) and (B2) correspond to $T=0.4T_c$; at this $\eta$, the zero of $\beta_3$ always occurs below this temperature. Panels (B3) and (B4) correspond to $T=0.5T_c$; at this $\eta$, the zero of $\beta_4+16Q^2/(\pi T)$ always occurs above this temperature. Panels (C1)--(C4) correspond to $\eta=0.5$. Panels (C1) and (C2) correspond to $T=0.3T_c$; at this $\eta$, the zero of $\beta_3$ always occurs below this temperature. Panels (C3) and (C4) correspond to $T=0.4T_c$; at this $\eta$, the zero of $\beta_4+16Q^2/(\pi T)$ always occurs above this temperature. Panels (D1)--(D4) correspond to $\eta=0.6$. Panels (D1) and (D2) correspond to $T=0.27T_c$; at this $\eta$, the zero of $\beta_3$ always occurs below this temperature. Panels (D3) and (D4) correspond to $T=0.3T_c$; at this $\eta$, the zero of $\beta_4+16Q^2/(\pi T)$ always occurs above this temperature. Panels (E1)--(E4) correspond to $\eta=0.7$. Panels (E1) and (E2) correspond to $T=0.23T_c$; at this $\eta$, the zero of $\beta_3$ always occurs below this temperature. Panels (E3) and (E4) correspond to $T=0.26T_c$; at this $\eta$, the zero of $\beta_4+16Q^2/(\pi T)$ always occurs above this temperature. Panels (F1)--(F4) correspond to $\eta=0.8$. Panels (F1) and (F2) correspond to $T=0.2T_c$; at this $\eta$, the zero of $\beta_3$ always occurs below this temperature. Panels (F3) and (F4) correspond to $T=0.22T_c$; at this $\eta$, the zero of $\beta_4+16Q^2/(\pi T)$ always occurs above this temperature. Panels (G1)--(G4) correspond to $\eta=0.88$. Panels (G1) and (G2) correspond to $T=0.18T_c$; at this $\eta$, the zero of $\beta_3$ always occurs below this temperature. Panels (G3) and (G4) correspond to $T=0.19T_c$; at this $\eta$, the zero of $\beta_4+16Q^2/(\pi T)$ always occurs above this temperature. }
\label{Fig12}
\end{figure}
%%%%%%%%%%%%%%%%%%%%%%%%%%%%%%%%%%%%%%%%%%%%%%%%%%%%%%%%%

%%%%
%%%%%%%%%%%%%%%%%%%%%%%%%
\begin{figure}[t]
\includegraphics[width=16cm]{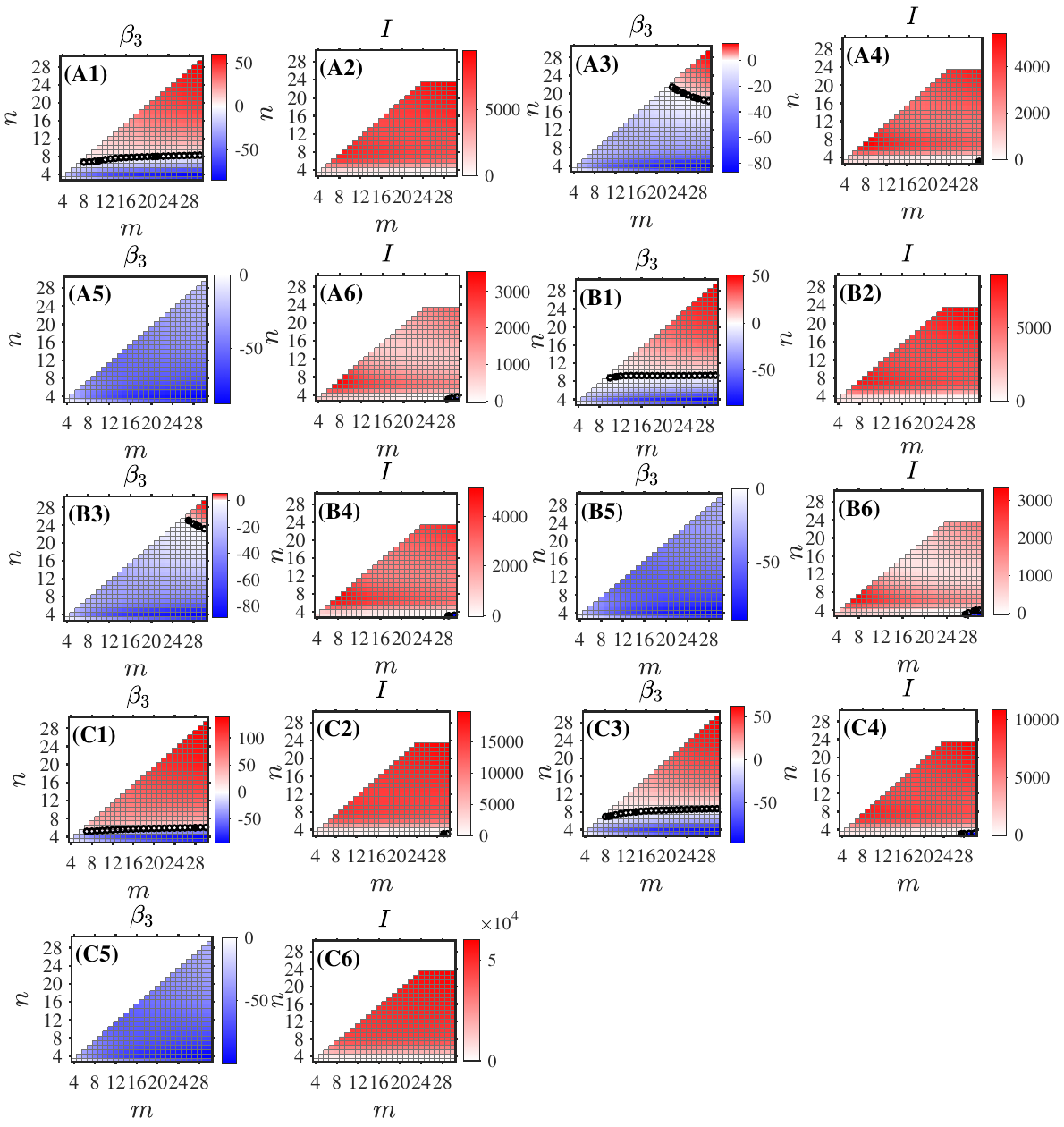}
\caption{Characteristic behaviors in the regime $0.89 \leq \eta \leq 1$ are shown; the behavior for $\eta=1$ is presented in the main text. In this regime, the zeros of both $\beta_3$ and $\beta_4+16Q^2/(\pi T)$ can occur, and they can occur at the same temperature. However, their zero loci in the $(m,n)$ plane do not intersect at any fixed temperature. Panels (A1)--(A6) correspond to $\eta=0.89$: panels (A1) and (A2) correspond to $T=0.17T_c$; at this $\eta$, the zero of $\beta_3$ always occurs below this temperature. Panels (A3) and (A4) correspond to $T=0.18T_c$. Panels (A5) and (A6) correspond to $T=0.19T_c$; at this $\eta$, the zero of $\beta_4+16Q^2/(\pi T)$ always occurs above this temperature. Panels (B1)--(B6) correspond to $\eta=0.90$: panels (B1) and (B2) correspond to $T=0.17T_c$; at this $\eta$, the zero of $\beta_3$ always occurs below this temperature. Panels (B3) and (B4) correspond to $T=0.18T_c$, and panels (B5) and (B6) correspond to $T=0.19T_c$; at this $\eta$, the zero of $\beta_4+16Q^2/(\pi T)$ always occurs above this temperature. Panels (C1)--(C6) correspond to $\eta=0.95$: panels (C1) and (C2) correspond to $T=0.15T_c$; at this $\eta$, the zero of $\beta_3$ always occurs below this temperature. Panels (C3) and (C4) correspond to $T=0.16T_c$; at this $\eta$, the zero of $\beta_4+16Q^2/(\pi T)$ always occurs above this temperature. Panel (C5) corresponds to $T=0.18T_c$, at which no zeros of $\beta_3$ occur, and panel (C6) corresponds to $T=0.13T_c$, at which no zeros of $\beta_4+16Q^2/(\pi T)$ occur.}
\label{Fig13}
\end{figure}
%%%%%%%%%%%%%%%%%%%%%%%%%%%%%%%%%%%%%%%%%%%%%%%%%%%%%%%%%

%%%%%%%%%%%%%%%%%%%%%%%%%%%%%%%%%%%%%%%%%%%%%%%%%%%%%%%%%